\documentclass{aa}
\usepackage{hyperref}
\usepackage{natbib} 
\usepackage{booktabs,graphicx,blindtext}
\usepackage{listings}
\usepackage{color} 

\begin{document}

\renewcommand{\arraystretch}{1.2}

\titlerunning{Automated Classification Pipeline for Periodic Variable Stars}
\title{A Package for the Automated Classification of Periodic Variable Stars}

\authorrunning{D.-W. Kim et al.}
\author{Dae-Won Kim\inst{1,}\thanks{current address: 
Institute of Astronomy and Astrophysics, Academia Sinica, PO Box 23-141, Taipei 10617, Taiwan}\and
Coryn A.L. Bailer-Jones\inst{1}}
\institute{
Max-Planck Institute for Astronomy, K\"{o}nigstuhl 17, D-69117 Heidelberg, Germany}

\abstract{We present a machine learning package for the classification of periodic variable stars.  Our package is intended to be general: it can classify any single band optical light curve
comprising at least a few tens of observations covering durations from weeks to years with arbitrary time sampling.
We use light curves of periodic variable stars taken from OGLE and EROS-2 to train the model.  To make our classifier relatively survey-independent, it is trained on 16 features extracted from the light curves (e.g., period, skewness, Fourier amplitude ratio). The model classifies light curves into one of seven superclasses -- $\delta$ Scuti, RR Lyrae, Cepheid, Type II Cepheid, eclipsing binary, long-period variable, non-variable -- as well as subclasses of these, such as ab, c, d, and e types for RR Lyraes.  When trained to give only superclasses, our model achieves 0.98 for both recall and precision as measured on an independent validation dataset (on a scale of 0 to 1).
When trained to give subclasses, it achieves 0.81 for both recall and precision.  The majority of misclassifications of the subclass model is caused by confusion within a superclass rather than between superclasses. 
To assess classification performance of the subclass model, we applied it to the MACHO, LINEAR, and ASAS periodic variables, which gave recall/precision of 0.92/0.98, 0.89/0.96, and 0.84/0.88, respectively.
We also applied the subclass model to Hipparcos periodic
variable stars of many other variability types that do not exist in our training set,
in order to examine how much those types degrade the classification performance
of our target classes.
In addition, we investigate how the performance varies with the number of data points and duration of observations. We find that recall and precision do not vary significantly if
there are more than 80 data points and the duration is more than a few weeks.
The classifier software of the subclass model is available (in \texttt{Python})
from the \href{https://goo.gl/xmFO6Q}{GitHub repository (https://goo.gl/xmFO6Q)}.
}


\maketitle

\section{Introduction}
\label{sec:introduction}

Periodic variable stars have played a central role in our learning about the Universe. RR Lyrae stars, for instance, have been used to determine distances to globular clusters, to study Galactic structure, and also to trace the history of the formation of our Galaxy \citep{Vivas2001ApJ...554L..33V, Carretta2000ApJ...533..215C, Catelan2009ApSS.320..261C, Sesar2011rrls.conf..135S}.  Cepheid variables are important standard candles for measuring distances to external galaxies \citep{Freedman2001ApJ...553...47F, Riess2011ApJ...730..119R}.  Eclipsing binaries are vital for estimating stellar masses \citep{Torres2010AARv..18...67T}, and $\delta$ Scuti are used in asteroseismological studies to learn about the internal structure of stars \citep{Brown1994ARAA..32...37B}.

As the volume of astronomical survey data grows, the reliable automated detection and classification of objects including periodic variable stars is becoming increasingly important. LSST, for instance, will produce about 20 TB of data per night, requiring robust and powerful automated methods to classify the sources and light curves (see \citealt{Ivezic2008arXiv0805.2366I} and references therein).  The Gaia satellite \citep{Perryman2001AA...369..339P} is scanning the whole sky to monitor a billion stars,
and is using machine learning and statistical analysis techniques to classify the sources and to estimate the astrophysical parameters (effective temperature, extinction, etc.) \citep{BailerJones2013AA...559A..74B}
as well as to analyze the light curves \citep{Eyer2014EAS....67...75E}.

In this paper, we introduce UPSILoN: A{\bf U}tomated Classification for {\bf P}eriodic Variable {\bf S}tars using Mach{\bf I}ne Lear{\bf N}ing. This software package classifies a light curve into a class of a periodic variable star in an automatic manner. It can be applied to light curves from optical surveys regardless of survey-specific characteristics (e.g., photometric accuracy, sampling rate, duration, etc.) as long as the light curves satisfy a few broad conditions:
\begin{itemize}
\item it contains at least a few tens of data points that sample the characteristic variability well;
\item the observation duration is more than a few weeks;
\item the light curve is obtained in an optical band.
\end{itemize}
If multiple optical bands are available, UPSILoN can classify them separately. It does not use -- or rely on -- color information. UPSILoN is currently trained to classify into six types of periodic variables: $\delta$ Scuti, RR Lyrae, Cepheid, Type II Cepheid, eclipsing binary, long-period variable, and their subclasses.  UPSILoN can also separate non-variables from these periodic variables, which is important because the majority of light curves from most time-series surveys are non-variables.

Some previous studies have investigated the possibility of producing general purpose classifiers such as UPSILoN.  For instance, \citet{Blomme2011MNRAS.418...96B} used a mixture of Hipparcos \citep{Perryman1997AA...323L..49P}, OGLE \citep{Udalski1997AcA....47..319U}, and CoRoT \citep{Auvergne2009AA...506..411A} sources to train a classification model, and then applied this to the TrES dataset \citep{ODonovan2009nsted.cat....6O}.  \citet{Debosscher2009AA...506..519D, Richards2012ApJ...744..192R} used Hipparcos and OGLE sources to train classification models and applied them to CoRoT and ASAS \citep{Pojmanski1997AcA....47..467P} data, respectively.  The latest work of this kind was done by \citet{Paegert2014AJ....148...31P} who developed a periodic variable classifier optimized for eclipsing binaries based on neural networks \citep{Mackay2003itil.book.....M}.  Most of these studies confirmed that their trained models successfully classified target sources from other surveys that were not included in their training sets.

UPSILoN differs from these previous works in several ways:

\begin{itemize}
	\item General purpose classifier
	
        UPSILoN has been trained to achieve
	the best classification quality for a broad set of periodic variables,
	rather than having been optimized for a specific type of variable.
	
	\vspace{0.1cm}	
	\item Rich training set
	
	We use a mixture of the OGLE
	and EROS-2 periodic variable stars to 
	provide a relatively complete set of periodic variables for training
	(see section \ref{sec:training_set}).
	
	\vspace{0.1cm}	
	\item Simple and robust variability features
	
	We employ features that have been shown to quantify
	variability of both periodic and non-periodic light curves 
	\citep{Kim2011ApJ...735...68K,Kim2012ApJ...747..107K,Kim2014AA...566A..43K,Long2012PASP..124..280L}.
	Most of these features are simple and easily calculated (see section \ref{sec:features}).

	\vspace{0.1cm}		
	\item Random forest \citep{Breiman2001}
	
	We use the random forest algorithm, 
	which is one of the most successful machine learning methods in
	many astronomical classification problems 
	(e.g., \citealt{Carliles2010ApJ...712..511C, Pichara2012MNRAS.427.1284P, 
	Masci2014AJ....148...21M, Nun2014ApJ...793...23N} and references therein).
	This is described in section \ref{sec:training_classification_model}.

	\vspace{0.1cm}		
	\item Application to MACHO, LINEAR, ASAS, and Hipparcos datasets
	
	To assess classification quality,
	we applied UPSILoN to the MACHO, LINEAR, ASAS, and Hipparcos 
	periodic variables, as described in 
	sections \ref{sec:application} and \ref{sec:application_2}.

	\vspace{0.1cm}		
	\item Experiments with resampled MACHO and ASAS light curves
	
	We resampled the MACHO and ASAS periodic variables' light curves in order to 
	build samples containing different number of measurements
	over different observation durations.
	Classification results using these samples are given in section \ref{sec:resampled_test}.

	\vspace{0.1cm}		
	\item Ready-to-use \texttt{Python} package
	
	The UPSILoN package including the random forest model
	is available from the \href{https://goo.gl/xmFO6Q}{GitHub repository (https://goo.gl/xmFO6Q)}.
	
\end{itemize}

In section \ref{sec:classification_model}
we introduce the training set and variability features 
that we used to train the UPSILoN classification model, and we explain the training process.
Section \ref{sec:application}, \ref{sec:application_2}, and \ref{sec:resampled_test}
summarize the classification performance
using the periodic-variable light curves collected from 
the MACHO \citep{Alcock1996AJ....111.1146A},
LINEAR \citep{Stokes2000Icar..148...21S},
ASAS \citep{Pojmanski1997AcA....47..467P},
and Hipparcos \citep{Perryman1997AA...323L..49P} surveys.
Section \ref{sec:summary} gives a summary.

\section{UPSILoN Classifier}
\label{sec:classification_model}

\subsection{Training Set}
\label{sec:training_set}

\renewcommand{\arraystretch}{1.2}
\begin{table}
\small
\begin{center}
\caption{Acronyms of the variable types \label{tab:acronym}}
\begin{tabular}{cc}
\hline\hline
Variable type & Acronym \\
\hline
$\delta$ Scuti star & DSCT \\
RR Lyrae & RRL \\
Cepheid & CEPH \\
Type II Cepheid & T2CEPH \\
Eclipsing binary & EB \\
Long-period variable & LPV \\
Non-variables & NonVar \\
\hline
\end{tabular}
\end{center}
\end{table}

\renewcommand{\arraystretch}{1.2}
\begin{table*}
\small
\begin{center}
\caption{The number of sources in each class in the training set\label{tab:training_set}}
\begin{tabular}{rrrr}
\hline\hline
Superclass & Subclass & Number & Note \\
\hline
DSCT & & 3209 & \\
RRL & &  & \\
		& ab & 19\,921 & \\
		& c & 4974 & \\
		& d & 1077 & \\
		& e & 1327 & \\
CEPH & & &  \\
		& F & 2981 & fundamental \\
		& 1O & 2043 & first overtone \\
		& Other & 443 & \\
EB &  &  & \\
		& EC & 1398 & contact \\
		& ED & 19\,075 &  detached \\
		& ESD &  7339 & semi-detached \\
LPV & & &  \\
		& Mira AGB C & 1361 &  carbon-rich \\
		& Mira AGB O & 783 &  oxygen-rich \\
		& OSARG AGB & 25\,284 &  \\
		& OSARG RGB & 29\,516 &  \\
		& SRV AGB C & 6062 & carbon-rich \\
		& SRV AGB O & 8780 & oxygen-rich \\
T2CEPH & & 300 & \\
NonVar & & 8050 &  \\
\hline
Total & & 143\,923 & \\
\end{tabular}
\end{center}
\end{table*}

As the quality of classification strongly depends on the training data, it is important to 
construct a rich and clean training set.
We used a mixture of periodic variable stars 
from the OGLE \citep{Udalski1997AcA....47..319U} and 
EROS-2 \citep{Tisserand2007AA...469..387T} Large Magellanic Cloud (LMC) databases.
We chose these because they provide the most complete and well-sampled set of
light curves in the optical for several types of periodic variable stars.
We identified the OGLE periodic variables from several literature sources:
\citet{Soszynski2008AcA....58..163S, Soszynski2008AcA....58..293S, 
Soszynski2009AcA....59..239S, Soszynski2009AcA....59....1S, 
Poleski2010AcA....60....1P, Graczyk2011AcA....61..103G}.
For the EROS-2 periodic variables,
we used the training set of the periodic variable stars
from \citet{Kim2014AA...566A..43K}.
As described in that paper, this data set was visually
inspected and cleaned.
From these source we complied a catalog of periodic variables 
comprising $\delta$ Scuti stars, RR Lyraes, Cepheids, Type II Cepheids,
eclipsing binaries, and long-period variables (these are the superclasses).
Table \ref{tab:acronym} gives the acronyms of each superclass.
We also compiled subclasses (e.g., RR Lyrae ab, c, d, and e) of each class.
We then added non-variable sources to the catalog, 
which is important because a given set of survey data that 
we want to classify will generally include many such sources.
For these, we selected 5000 EROS-2 non-variables from \citet{Kim2014AA...566A..43K} and 5000 OGLE light curves randomly selected around the Large Magellanic Cloud using the website \href{http://ogledb.astrouw.edu.pl/~ogle/photdb/}{http://ogledb.astrouw.edu.pl/$\sim$ogle/photdb/} \citep{Szymanski2005AcA....55...43S}. 
We visually examined these light curves and removed light curves showing variability.
Note, however, that it remains difficult for any general classifier to identify non-variables on account of the wide range of survey-dependent characteristics of such stars (e.g., due to systematic trends, variable noise levels).  Thus although we added the OGLE and EROS-2 non-variable sources to our training set, this does not ensure that the classifier will efficiently exclude non-variables during the classification of light curves from other surveys.

Before extracting variability features (section \ref{sec:features}) 
from these training-set light curves, we first refined each light curve by

\begin{itemize}

\item removing data points with bogus photometry (e.g., 99.999 in EROS-2).
The EROS-2 light curves have many such measurements.

\item removing data points that lie more than 3$\sigma$ above or below the mean (no iteration), where $\sigma$ is the standard deviation of the light curve.
This cut does not significantly alter light curves
because it removes generally less than 1\% of the data points per light curve.

\end{itemize}

{\noindent}In order to build the cleanest possible training set, 
we then excluded light curves from the training set if

\begin{itemize}

\item the number of measurements was fewer than 100.

\item an estimated period was spurious, related to
the mean sidereal day ($\sim$0.997 days), 
the mean sidereal month ($\sim$27.322 days),
the mean synodic month ($\sim$29.531 days), and integer multiples thereof.
We determined additional spurious periods by examining 
a period versus period signal-to-noise ratio (SNR) plot and 
a period histogram as done by \citet{Kim2014AA...566A..43K}.
Most of the additional spurious periods are close to the sidereal day.

\end{itemize}

Table \ref{tab:training_set} shows the number of each 
variable type in the final training set.
``CEPH Other'' class contains subclasses of Cepheid variables
such as second overtone (2O),
double-mode in fundamental and first overtone (F/1O), 
double-mode in first and second overtone (1O/2O), etc.
The total number of sources in the final training set is 143\,923.
As the table shows, the training set contains the major and strongly
varying periodic variables. The goal of UPSILoN is not to classify
other variability types such as rotational variability, eruptive variability, etc.
Section \ref{sec:application_2} presents 
the possible contamination caused by such variability types 
using a set of Hipparcos periodic variable stars from \citet{Dubath2011MNRAS.414.2602D}.

The average number of data points in either an OGLE
or an EROS-2 light curve is about 500.
The observation duration is about seven years for the EROS-2 light curves,
and about eight years for the OGLE light curves.

\subsection{Survey-Independent Variability Features}
\label{sec:features}

\renewcommand{\arraystretch}{1.2}
\begin{table*}
\small
\begin{center}
\caption{16 Variability features \label{tab:variability_features}}
\begin{tabular}{clc}
\hline\hline
Feature & Description & Reference$^a$ \\
\hline
Period & Period derived by the Lomb-Scargle algorithm \\
$\psi^{\eta}$ & $\eta$ \citep{Neumann1941} of a phase-folded light curve & \\
$\psi^{CS}$ & Cumulative sum index of a phase-folded light curve \\
$R_{21}$ & 2$^{nd}$ to 1$^{st}$ amplitude ratio derived using the Fourier decomposition \\
$R_{31}$ & 3$^{rd}$ to 1$^{st}$ amplitude ratio derived using the Fourier decomposition \\
$\phi_{21}$ & Difference between 2$^{nd}$ and 1$^{st}$ phase  derived using the Fourier decomposition \\
$\phi_{31}$ & Difference between 3$^{rd}$ and 1$^{st}$ phase derived using the Fourier decomposition \\
$\gamma_1$  & Skewness \\
$\gamma_2$ & Kurtosis \\
$K$ & Stetson $K$ index \\
$Q_{3-1}$ & 3$^{rd}$ quartile (75\%) - 1$^{st}$ quartile (25\%) \\
$A$ & A ratio of magnitudes brighter or fainter than the average & In this work \\ 
$H_1$ & Amplitude derived using the Fourier decomposition & In this work \\
$W$ & Shapiro--Wilk normality test statistics & In this work \\
$m_{p10}$ & 10\% percentile of slopes of a phase-folded light curve & \citet{Long2012PASP..124..280L} \\
$m_{p90}$ & 90\% percentile of slopes of a phase-folded light curve & \citet{Long2012PASP..124..280L} \\
\hline
\end{tabular}
\end{center}
$^a$ From \citet{Kim2014AA...566A..43K} if not specified.
\end{table*}

We used a total of 16 variability features (Table \ref{tab:variability_features}) to quantify variability of a light curve.  Note that we only chose features that are relatively survey-independent\footnote{The minimum requirements for a light curve are given in section \ref{sec:introduction}.} and excluded features that are only available in certain surveys (e.g., ones based on colors).  This is important because our goal is to develop a general purpose classifier that can be applied to an essentially any long-duration time-series survey done in an optical band.  Among these 16 features, 11 are taken from our previous work \citep{Kim2011ApJ...735...68K,Kim2012ApJ...747..107K,Kim2014AA...566A..43K}, and two are from \citet{Long2012PASP..124..280L}.  These publications show that these features are useful and robust to classify either periodic or non-periodic variables.  We added three more features, $A$, $H_1$, and $W$, defined as follows.

\begin{itemize}

\item $A$

The ratio of the squared sum of residuals
of magnitudes that are either brighter than or fainter than the mean magnitude.
This is defined as
\begin{alignat}{2}
A &= \frac{\sigma_f}{\sigma_b}~~~\text{where} \\
\sigma_f^2 &= \frac{1}{N} \sum_i^N {(m_i - \overline{m})}^2 \ , \\
\sigma_b^2 &= \frac{1}{M} \sum_j^M {(m_j - \overline{m})}^2 \ ,
\end{alignat}
$\overline{m}$ is the mean magnitude of a light curve,
$m_i$ are measurements fainter than $\overline{m}$,
and $m_j$ are measurements brighter than $\overline{m}$.
If a light curve has a sinusoidal-like variability,
$A$ is close to 1. For EB-like variability, 
having sharp flux gradients around its eclipses,
$A$ is larger than 1.

\item $H_1$

The RMS amplitude of the first Fourier component of the Fourier decomposition of the light curve
\citep{Petersen1986AA...170...59P}.
This is
\begin{equation}
H_1 = \sqrt{a_1^2 + b_1^2} 
\end{equation}
with the Fourier series defined as 
\begin{equation}
m_0 + \sum_{k=1}^M [  a_k \sin(2 \pi k f) + b_k \cos(2 \pi k f) ]
\end{equation}
where $M$ is the order of the Fourier series
and $f$ is the phase: $0 \leq f < 1; j = 1, 2, \ldots, N$, where $N$ is
the number of data points. $m_0$, $a_k$ and $b_k$ are  parameters to be fitted.
In this work, the value of $M$ is 5.
\item $W$

The Shapiro--Wilk statistic, which tests the null hypothesis that the 
samples (e.g., measurements in the light curve) are from a normal distribution \citep{shapiro1965analysis}.
The Shapiro--Wilk test is generally more powerful
than other normality tests (e.g., Kolmogorov-Smirnov test, Anderson-Darling test, etc.)
for different sample sizes (see \citealt{Saculinggan12013} and references therein).

\end{itemize}

\begin{figure}
\begin{center}
       \includegraphics[width=0.5\textwidth]{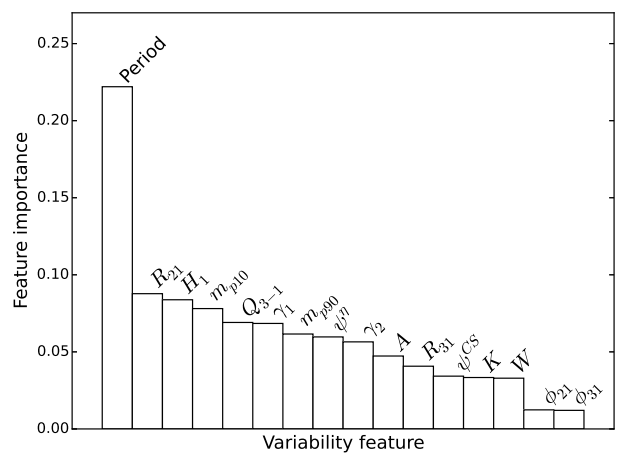}
\end{center}
    \caption{Feature importance as estimated using the random forest algorithm.
    As expected, period is the most powerful feature for separating periodic variables, although the other 15 features
    %(except perhaps the last two) 
    are non unimportant.}
    \label{fig:feature_importance}
\end{figure}

Fig. \ref{fig:feature_importance} shows the importance of each feature in the trained subclass classifier 
(see section \ref{sec:training_classification_model}) as estimated by the random forest algorithm on the training data.  A random forest ranks the importance of $i^{th}$ feature by using {\em{out-of-bag}} (oob) samples, which is about one-third of the training data 
and not used for constructing a tree. A random forest predicts classes of the oob samples
and counts the votes for the correct classes. A random forest then randomly permutes
the values of $i^{th}$ feature in the oob samples, predicts classes again with the permuted oob samples, 
and counts the correct votes. The average difference of the count over all trees is the importance of $i^{th}$ feature.
In addition, a random forest can also estimate the feature importance
based on the mean decrease in node impurity (e.g., Gini impurity),
which is the total sum of the decrease in node impurity averaged over all trees.
This method gives generally consistent results with the method using permutation 
(see \citealt{breiman1984classification, Breiman2001} and references therein).
In this work, we used the Gini impurity measure to deduce the feature importance.

%2015UPSILoN/plot_figures/feature_importance.py

To extract variability features from the training-set light curves,
we used EROS-2 blue-band, $B_E$, light curves.
For OGLE, we used $I$-band light curves.
EROS-2 $B_E$-band and OGLE $I$-band light curves generally have
more data points and/or better photometric accuracy
than EROS-2 red-band, $R_E$, and OGLE $V$-band light curves, respectively.

\subsection{Training a Classification Model}
\label{sec:training_classification_model}

\renewcommand{\arraystretch}{1.2}
\begin{table*}
\small
\begin{center}
\caption{Classification quality of the trained model with subclasses\label{tab:model_with_subclasses}}
\begin{tabular}{ccccc}
\hline\hline
Superclass & Subclass & Precision & Recall &$F_1$ \\
\hline
DSCT &  & 0.93 & 0.89 & 0.91 \\
RRL & &  &  & \\
		& ab & 0.99  & 0.99 & 0.99 \\
		& c & 0.92 & 0.96 & 0.94 \\
		& d & 0.94 & 0.83 & 0.88 \\
		& e & 0.90 & 0.91 & 0.91 \\
CEPH & & &   & \\
		& F & 0.98 & 0.98  & 0.98 \\
		& 1O & 0.93 & 0.92  & 0.92 \\
		& Other & 0.88 & 0.76 & 0.81 \\
EB &  &  &  & \\
		& EC & 0.76 & 0.68  & 0.72 \\
		& ED & 0.90 & 0.91  & 0.90 \\
		& ESD &  0.80 & 0.74  & 0.77 \\
LPV & & &   & \\
		& Mira AGB C & 0.83 & 0.91 & 0.87 \\
		& Mira AGB O & 0.76 & 0.74   & 0.75 \\
		& OSARG AGB & 0.66 & 0.65  & 0.65 \\
		& OSARG RGB & 0.74 & 0.80  & 0.77 \\
		& SRV AGB C & 0.70 & 0.60  & 0.65 \\
		& SRV AGB O & 0.70 & 0.67  & 0.69 \\
T2CEPH & & 0.91 & 0.49 & 0.64 \\
NonVar & & 0.87 & 0.89  & 0.88 \\
\hline
Average$^a$ & & 0.81 & 0.81 & 0.81 \\
\end{tabular}
\end{center}
$^a$ Weighted average by the number of source in a class
\end{table*}

A random forest classifier \citep{Hastie09, Breiman2001} 
is based on an ensemble of decision trees \citep{Quinlan1993}.
Each decision tree is trained using a randomly selected subset of the features at each node.
The random forest uses a majority voting strategy to assign 
probabilities to all labels (here the variability classes)
for a given sample (here the light curve). 
We accept the variability class corresponding to the highest probability
among those assigned probabilities.
See \citet{Kim2014AA...566A..43K} for a more detailed description of
the random forest algorithm.

\begin{figure}
\begin{center}
       \includegraphics[width=0.5\textwidth]{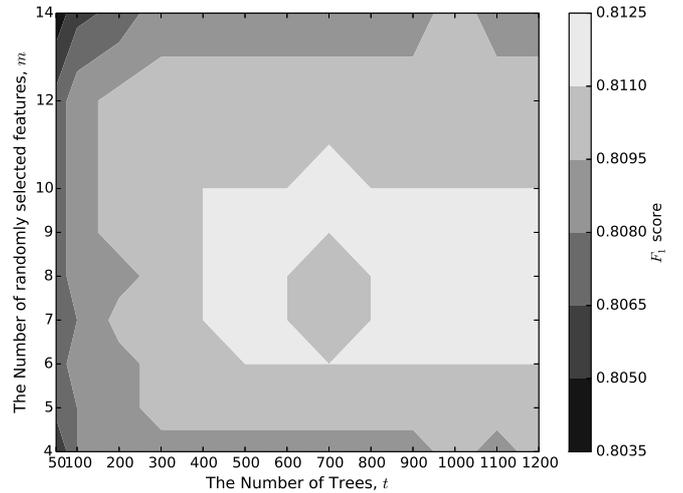}
\end{center}
    \caption{$F_1$ scores of the trained random forest models
    with different $t$ and $m$. Note that the $F_1$ scores are quite uniform
    over the parameter space, showing differences of less than 0.01.}
    \label{fig:grid_search}
\end{figure}

To train a random forest model, two hyperparameters
must be set. These are the number of trees, $t$,
and the number of randomly selected features at each node of trees, $m$.
Each tree is fully grown without pruning.
Before tuning the parameters,
we split the training set into two subsets, $T_1$ and $T_2$,
each of which contains 50\% of each class in the training set. 
We used $T_1$ to optimize the hyperparameters
by performing a brute-force search over a two-dimensional grid in $t$ and $m$, using ten-fold cross-validation.
The metric for model performance is the $F_1$ score, which
is the harmonic mean of the precision and recall, i.e.,
\begin{alignat}{2}
F_1 &= 2 \times \displaystyle\frac{\text{precision} \times \text{recall}}{\text{precision + recall}}~~~\text {where } \\
\text{precision} &= \displaystyle\frac{\text{True Positive}}{\text{True Positive + False Positive}} \ , \\
\text{recall} &= \displaystyle\frac{\text{True Positive}}{\text{True Positive + False Negative}} \ , \label{eqn:recall}
\end{alignat}
and $F_1$ varies between 0 and 1.
This is computed for each class separately, and the individual scores then averaged weighted by the number of objects in that class.
Fig. \ref{fig:grid_search} shows the results of the grid search.  
The difference between the best and worst models is 0.009, 
so the exact choice of $t$ and $m$ is unimportant.
The classification uncertainty derived during the ten-fold cross-validation is $\pm$0.005, 
which is comparable to this difference.

\begin{figure*}
\begin{center}
       \includegraphics[width=0.9\textwidth]{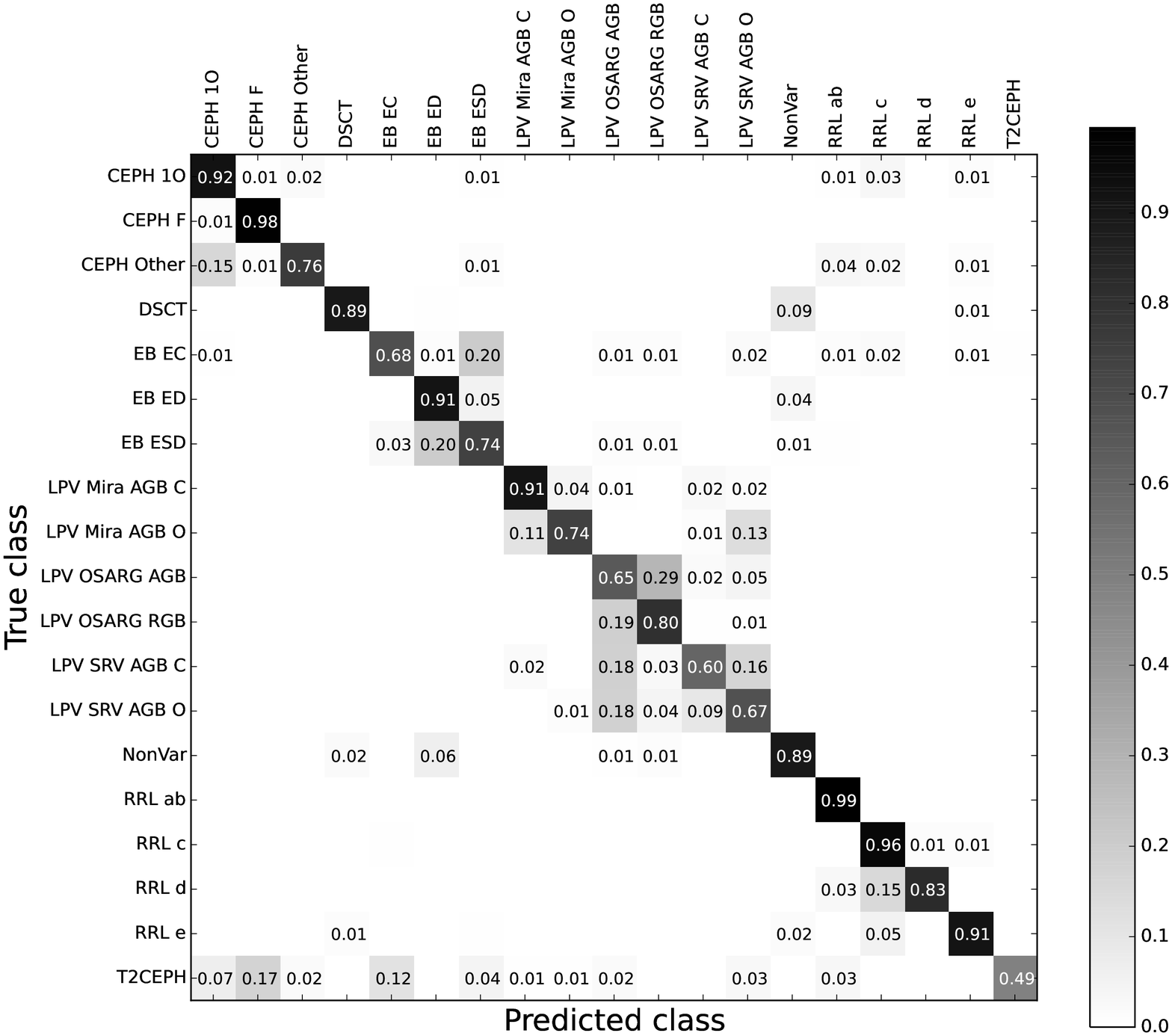}
\end{center}
    \caption{Confusion matrix of the subclass model.
   Each cell shows the fraction of objects of that true class (rows) assigned to the predicted classes (columns) on a gray scale. Thus the values on the leading diagonal are the recall rate (equation \ref{eqn:recall}). We show the number only if it is larger than or equal to 0.01.
See Table \ref{tab:model_with_subclasses} for the numerical values of the recall
    and precision for each class.}
    \label{fig:model_map_subclass}
\end{figure*}

With $t=700$ and $m=10$ (the best values) we trained a classification model using the subset $T_1$
and applied it to $T_2$ in order to 
assess the classification quality. This is summarized in 
Table \ref{tab:model_with_subclasses}.
Recall and precision for DSCT, RRL, CEPH, and their subclasses are
generally higher than 0.90, sometimes as high as 0.99.
The majority of misclassification is caused by confusion within a superclass, 
especially within EB and LPV, as can be seen in Fig. \ref{fig:model_map_subclass}.
In addition, if we train\footnote{The hyperparameter optimization is done separately.} 
a classification model without the subclasses, 
the $F_1$ score increases to 0.98 as shown in Table \ref{tab:model_without_subclasses}.

\renewcommand{\arraystretch}{1.2}
\begin{table}
\small
\begin{center}
\caption{Classification quality of the trained model without subclasses \label{tab:model_without_subclasses}}
\begin{tabular}{cccc}
\hline\hline
Superclass & Precision & Recall & $F_1$ \\
\hline
DSCT & 0.94 & 0.87 &  0.90\\
RRL &  0.99 & 0.99 & 0.99 \\
CEPH & 0.98 & 0.96& 0.97 \\
EB & 0.97 & 0.96 & 0.96 \\
LPV & 0.99 & 1.00 & 1.00 \\
T2CEPH & 0.90 & 0.48 & 0.62 \\
NonVar & 0.87 & 0.87 & 0.87 \\
\hline
Average$^a$ & 0.98 & 0.98 & 0.98\\
\end{tabular}
\end{center}
$^a$ Weighted average by the number of source in a class
\end{table}

Finally, we trained a classifier using all 16 variability features,
the entire training set (i.e., $T_1 + T_2$), and the subclasses
(again with $t = 700$ and $m = 10$).
Note that this model is the UPSILoN classifier that
we use to classify the MACHO, LINEAR, ASAS, and Hipparcos periodic variables
presented in the following sections.

\section{Validation of the Classifier on the MACHO, LINEAR, and ASAS Datasets}
\label{sec:application}

We applied the UPSILoN classifier to light curves found as periodic variables from three time-series surveys:
MACHO \citep{Alcock1996ApJ...461...84A};
ASAS \citep{Pojmanski1997AcA....47..467P};
LINEAR \citep{Palaversa2013AJ....146..101P}.
These each have different survey characteristics,
such as the number of data points, observation duration,
limiting magnitude, etc, so represent a suitable datasets to
assess the UPSILoN classifier's generalization performance.
Before we applied the classifier to these three datasets, 
we refined each light curve by 3$\sigma$ clipping,
and by excluding light curves having spurious periods,
as described in section \ref{sec:classification_model}.
We then extracted the 16 variability features from each light curve and used
UPSILoN to predict a variability class.

\subsection{MACHO}
\label{sec:MACHO}

\renewcommand{\arraystretch}{1.2}
\begin{table}
\small
\begin{center}
\caption{The number of objects per true class in the MACHO dataset \label{tab:n_class_macho}}
\begin{tabular}{rr}
\hline\hline
Class & Number \\
\hline
RRL & 270\\
CEPH & 75\\
EB & 189\\
LPV & 336\\
\hline
Total & 870\\
\end{tabular}
\end{center}
\end{table}

\begin{table*}
\centering
\caption{Classification quality of the MACHO dataset. \label{tab:MACHO_testset}
Left: the UPSILoN classifier. Center: random classifier with a training set prior. Right:
random classifier with a uniform prior.}
\minipage{0.32\textwidth}
\resizebox{\linewidth}{!}{%
\begin{tabular}{cccc}
\hline\hline
Superclass & Precision & Recall & $F_1$ \\
\hline
RRL & 1.00 & 0.85 & 0.92 \\
CEPH & 1.00 & 0.75 & 0.85 \\
EB & 0.90 & 0.97 & 0.94 \\
LPV & 0.99 & 0.98 & 0.99 \\
\hline
Average & 0.98 & 0.92 & 0.94 \\
\end{tabular}}
\endminipage\hfill%
\minipage{0.32\textwidth}
\resizebox{\linewidth}{!}{%
\begin{tabular}{cccc}
\hline\hline
Superclass & Precision & Recall & $F_1$ \\
\hline
RRL & 0.27 & 0.16 & 0.20 \\
CEPH & 0.03 & 0.01 & 0.02 \\
EB & 0.21 & 0.20 & 0.20 \\
LPV & 0.39 & 0.49 & 0.44 \\
\hline
Average & 0.28 & 0.28 & 0.28 \\
\end{tabular}}
\endminipage\hfill%
\minipage{0.32\textwidth}
\resizebox{\linewidth}{!}{%
\begin{tabular}{cccc}
\hline\hline
Superclass & Precision & Recall & $F_1$ \\
\hline
RRL & 0.34 & 0.24 & 0.28 \\
CEPH & 0.13 & 0.24 & 0.17 \\
EB & 0.24 & 0.16 & 0.20 \\
LPV & 0.35 & 0.29 & 0.32 \\
\hline
Average & 0.31 & 0.24 & 0.27 \\
\end{tabular}}
\endminipage
\end{table*}

\begin{figure*}
\begin{center}
        \includegraphics[width=1.0\textwidth]{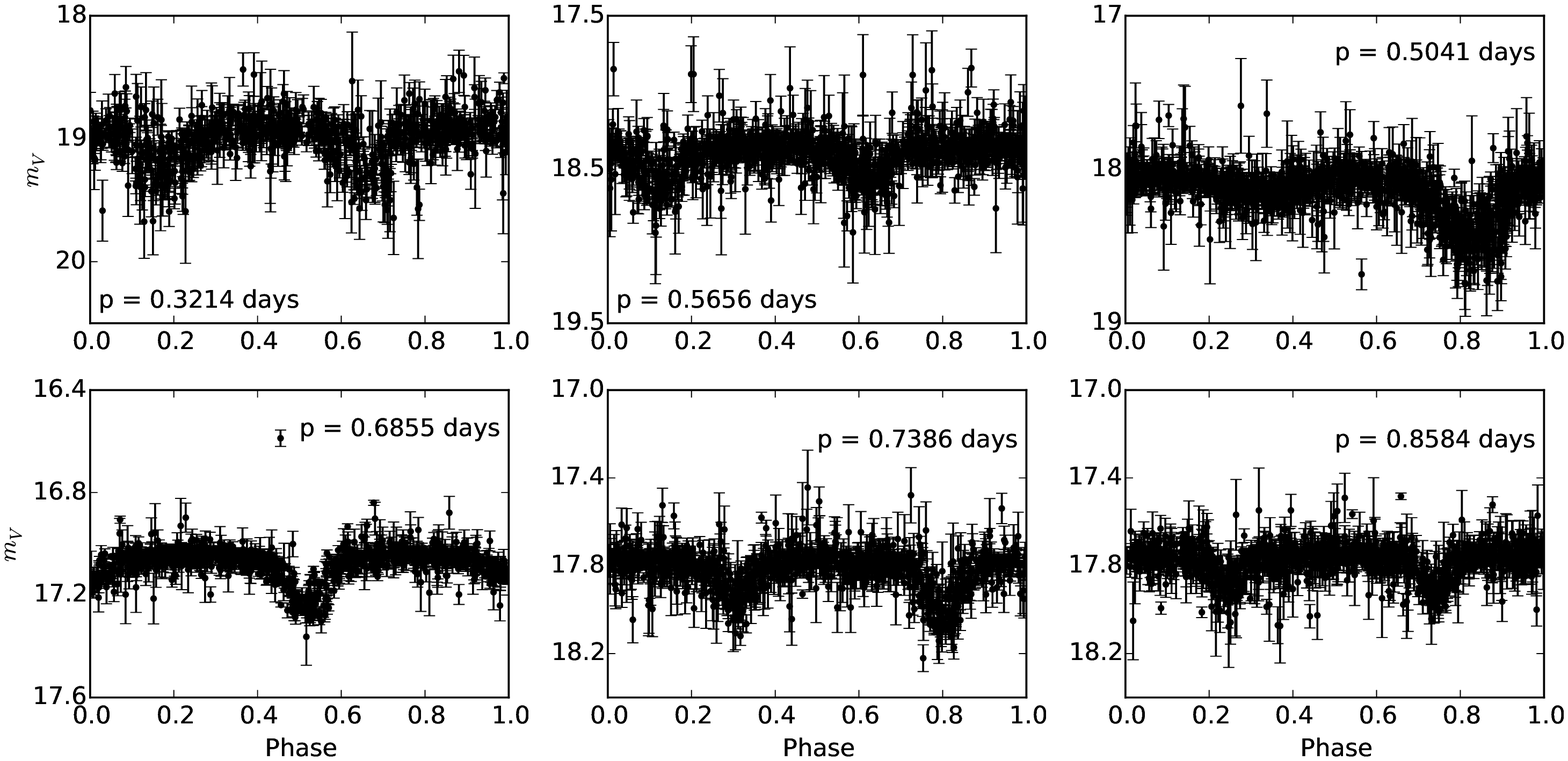}
\end{center}
    \caption
           {Phased-folded light curves of the MACHO RRLs classified as
           EBs by the UPSILoN classifier. 
           They show EB-like variability, implying
           that the MACHO classification might be incorrect.
           $m_V$ is the Johnson V-band magnitude derived from the
           MACHO magnitude using the conversion equation from
            \citet{Kunder2008AJ....136.2441K}.
            In each panel we show the periods of the light curves obtained
            from the Lomb-Scargle algorithm.
            We used twice these periods to plot the phase-folded light curves,
            because Lomb-Scargle
            occasionally returns half of the true period for EB-like variability.}
    \label{fig:MACHO_RRL_EB}
\end{figure*}

We obtained MACHO $B$-band light curves of
periodic variables  from \citet{Kim2011ApJ...735...68K}.
The total number of light curves is 870 (Table \ref{tab:n_class_macho}).
The number of data points in each light curve is between 200 and 1500, 
while 50\% of the light curves have more than 1000 data points. 
The duration of the light curves (i.e., observation duration) is about 7 years.
The limiting magnitude of the MACHO survey is $V \sim 21$.

In the left panel of Table \ref{tab:MACHO_testset} we show the classification performance
achieved by UPSILoN.
MACHO does not provide subclass classifications so
thus we merged predicted subclasses into corresponding superclasses.
As the table shows, the average $F_1$ score is 0.94, which confirms that
the UPSILoN classifier is successful at classifying the MACHO periodic variables.
Nevertheless, the recall of CEPH is only 0.75, meaning that
25\% of CEPH were classified as another variability type.
Almost all of these were
classified as T2CEPH, which is not surprising since
CEPH and T2CEPH are similar.
Note that MACHO did not use the class T2CEPH,
so it is quite possible that some of these sources are in fact of class T2CEPH.
The recall for RRL is 0.85, i.e., 15\% of true RRLs were misclassified.
Many of these were classified as EB,
and we visually confirmed that they indeed show EB-like variability 
(Fig. \ref{fig:MACHO_RRL_EB}).

Table \ref{tab:MACHO_testset} also shows the classification performance using a random classifier with a training-set prior (middle panel), and an uniform prior (right panel).  The term ``random classifier'' means that we randomly assigned a class to each MACHO light curve.  The probability of assignment to a given class is proportional to the frequency of the class in the training set in the middle panel, and is equal for the classes in the right panel.  We do these tests to get a measure of the minimum expected performance (which is not zero), and to quantify to what degree beyond ``random'' the data actually inform about the class.  As expected, the precision and recall of both random classifiers are always considerably lower than that achieved by UPSILoN.

\subsection{LINEAR}

\renewcommand{\arraystretch}{1.2}
\begin{table}
\small
\begin{center}
\caption{Number of objects per true class in the LINEAR dataset \label{tab:n_class_linear}}
\begin{tabular}{rrr}
\hline\hline
Superclass & Subclass & Number \\
\hline
DSCT & & 104\\
RRL & & \\
	& ab & 2415\\
	& c & 740\\
EB & &1965\\
LPV && 55\\
\hline
Total & & 5279\\
\end{tabular}
\end{center}
\end{table}

\begin{table*}
\centering
\caption{Classification quality of the LINEAR dataset. \label{tab:LINEAR_testset}
Left: the UPSILoN classifier. Center: random classifier with a training set prior. Right:
random classifier with a uniform prior.}
\minipage{0.32\textwidth}
\resizebox{\linewidth}{!}{%
\begin{tabular}{ccccc}
\hline\hline
Superclass & Subclass & Precision & Recall &$F_1$ \\
\hline
DSCT &  & 0.33 & 0.99 & 0.50 \\
RRL & &  &  & \\
		& ab & 0.99  & 0.91 & 0.95 \\
		& c & 0.92 & 0.81 & 0.86 \\
EB &  & 0.97  & 0.88  & 0.93 \\
LPV & & 0.95 & 1.00  & 0.97 \\
\hline
Average & & 0.96 & 0.89 & 0.92 \\
\end{tabular}}
\endminipage\hfill%
\minipage{0.32\textwidth}
\resizebox{\linewidth}{!}{%
\begin{tabular}{ccccc}
\hline\hline
Superclass & Subclass & Precision & Recall &$F_1$ \\
\hline
DSCT &  & 0.03 & 0.03 & 0.03 \\
RRL & &  &  & \\
		& ab & 0.48  & 0.14 & 0.21 \\
		& c & 0.15 & 0.04 & 0.07 \\
EB &  & 0.38  & 0.20  & 0.26 \\
LPV & & 0.01 & 0.49  & 0.02 \\
\hline
Average & & 0.38 & 0.15 & 0.20 \\
\end{tabular}}
\endminipage\hfill%
\minipage{0.32\textwidth}
\resizebox{\linewidth}{!}{%
\begin{tabular}{ccccc}
\hline\hline
Superclass & Subclass & Precision & Recall &$F_1$ \\
\hline
DSCT &  & 0.00 & 0.01 & 0.01 \\
RRL & &  &  & \\
		& ab & 0.48  & 0.06 & 0.10 \\
		& c & 0.15 & 0.05 & 0.08 \\
EB &  & 0.37  & 0.16  & 0.22 \\
LPV & & 0.01 & 0.29  & 0.02 \\
\hline
Average & & 0.38 & 0.10 & 0.14 \\
\end{tabular}}
\endminipage
\end{table*}

The light curves and catalog of LINEAR periodic variables
are from \citet{Palaversa2013AJ....146..101P}.
We selected sources with the classification uncertainty flags of 1 
and excluded others on the ground that their class assignment was ambiguous.
The total number of sources after the selection is
5279 (Table \ref{tab:n_class_linear}).
The number of data points of each light curve lies between 100 and 600, although
 most of the light curves have fewer than 300 data points.
The duration of the light curves is about 5 years
and the faintest magnitude among these periodic variables is r $\sim$ 17.

\begin{figure}
\begin{center}
        \includegraphics[width=0.5\textwidth]{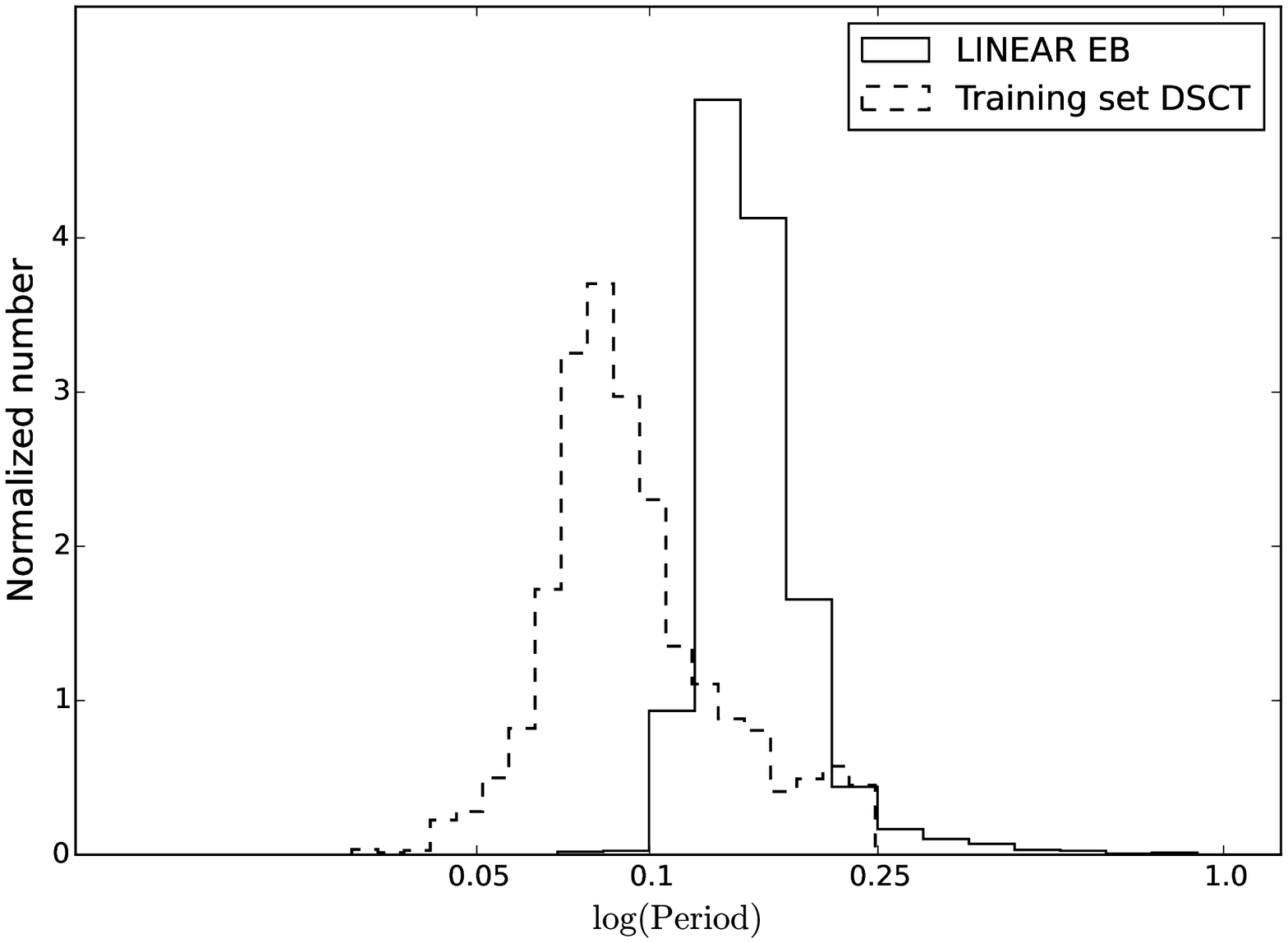}
\end{center}
    \caption
           {Period histograms of EB (solid line) from the LINEAR dataset.
           LINEAR EB have relatively shorter periods,
           which could cause misclassification of them into the
           short-period variables such as DSCT. 
           The dashed line is a period histogram of 
           the training set DSCT.}
    \label{fig:EB_period_histogram}
\end{figure}

The left panel of Table \ref{tab:LINEAR_testset} shows
the UPSILoN classification performance.
All classes except DSCT are successfully classified by UPSILoN.
The low precision for DSCT occurs because
199 of the 1965 EBs were misclassified as DSCT, of which there are only 104 true ones.
The reason for this is because 96\% of the estimated periods
of the LINEAR EBs are shorter than 0.25 days (Fig. \ref{fig:EB_period_histogram}),
which is associated with the problem that the Lomb-Scargle algorithm typically
returns half of the true period for EB-type variables in which
the shapes of the primary and secondary eclipses are similar
(see \citealt{Graham2013MNRAS.434.3423G} and references therein).
Consequently, almost all of the LINEAR EB periods are within
the period range for the DSCTs in the training set (i.e., $\leq$ 0.25 days), 
which could cause the misclassification.
Despite the overlapping ranges of
the estimated periods between LINEAR EB and DSCT,
88\% of EBs (i.e., 1\,734 among 1\,965) were correctly classified.

\subsection{ASAS}

\renewcommand{\arraystretch}{1.2}
\begin{table}
\small
\begin{center}
\caption{Number of objects per true class in the ASAS dataset \label{tab:n_class_asas}}
\begin{tabular}{rrr}
\hline\hline
Superclass & Subclass & Number \\
\hline
DSCT & & 520\\
RRL & & \\
	& ab & 1218\\
	& c & 305\\
CEPH && \\
	& F & 567\\
	& 1O & 179\\
EB & &\\
	& EC & 2550\\
	& ED & 2167\\
	& ESD & 823\\
LPV && 2450\\
\hline
Total & & 10\,779\\
\end{tabular}
\end{center}
\end{table}

\begin{table*}
\centering
\caption{Classification quality of the ASAS dataset. \label{tab:ASAS_testset}
Left: the UPSILoN classifier. Center: random classifier with a training set prior. Right:
random classifier with a uniform prior.}
\minipage{0.32\textwidth}
\resizebox{\linewidth}{!}{%
\begin{tabular}{ccccc}
\hline\hline
Superclass & Subclass & Precision & Recall &$F_1$ \\
\hline
DSCT &  & 0.70 & 0.90 & 0.79 \\
RRL & &  &  & \\
		& ab & 0.92  & 0.94 & 0.93 \\
		& c & 0.50 & 0.80 & 0.61 \\
CEPH & & &   & \\
		& F & 0.94 & 0.54  & 0.69 \\
		& 1O & 0.78 & 0.53  & 0.63 \\
EB &  &  &  & \\
		& EC & 0.91 & 0.67  & 0.77 \\
		& ED & 0.93 & 0.96  & 0.94 \\
		& ESD &  0.57 & 0.72  & 0.64 \\
LPV & & 0.96 & 0.96  & 0.96 \\
\hline
Average & & 0.88 & 0.84 & 0.85 \\
\end{tabular}}
\endminipage\hfill%
\minipage{0.32\textwidth}
\resizebox{\linewidth}{!}{%
\begin{tabular}{ccccc}
\hline\hline
Superclass & Subclass & Precision & Recall &$F_1$ \\
\hline
DSCT &  & 0.04 & 0.02 & 0.02 \\
RRL & &  &  & \\
		& ab & 0.11  & 0.13 & 0.12 \\
		& c & 0.02 & 0.03 & 0.03 \\
CEPH & & &   & \\
		& F & 0.04 & 0.02  & 0.02 \\
		& 1O & 0.01 & 0.01  & 0.01 \\
EB &  &  &  & \\
		& EC & 0.26 & 0.01  & 0.02 \\
		& ED & 0.21 & 0.14  & 0.17 \\
		& ESD &  0.07 & 0.04  & 0.05 \\
LPV & & 0.23 & 0.50  & 0.32 \\
\hline
Average & & 0.18 & 0.17 & 0.13 \\
\end{tabular}}
\endminipage\hfill%
\minipage{0.32\textwidth}
\resizebox{\linewidth}{!}{%
\begin{tabular}{ccccc}
\hline\hline
Superclass & Subclass & Precision & Recall &$F_1$ \\
\hline
DSCT &  & 0.06 & 0.07 & 0.06 \\
RRL & &  &  & \\
		& ab & 0.11  & 0.05 & 0.07 \\
		& c & 0.04 & 0.07 & 0.05 \\
CEPH & & &   & \\
		& F & 0.06 & 0.06  & 0.06 \\
		& 1O & 0.01 & 0.04  & 0.02 \\
EB &  &  &  & \\
		& EC & 0.22 & 0.06  & 0.09 \\
		& ED & 0.18 & 0.05  & 0.08 \\
		& ESD &  0.09 & 0.05  & 0.07 \\
LPV & & 0.23 & 0.32  & 0.26 \\
\hline
Average & & 0.17 & 0.11 & 0.12 \\
\end{tabular}}
\endminipage
\end{table*}

\begin{figure}
\begin{center}
        \includegraphics[width=0.5\textwidth]{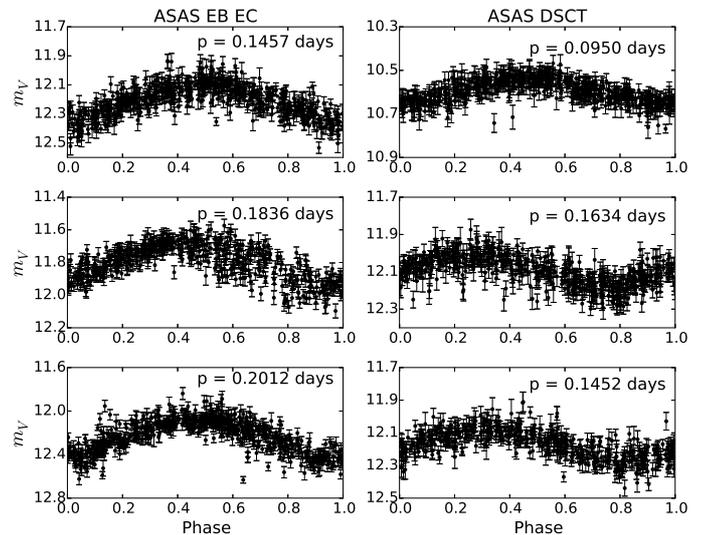}
\end{center}
    \caption
           {Phase-folded ASAS light curves of EB EC (left) and DSCT (right)
           from the ACVS classification. As the figure shows,
           there are no clear variability differences between them,
           which could yield misclassification.
           Periods (in days) of the light curves are given in each panel.}
    \label{fig:ASAS_EB_DSCT}
\end{figure}

We downloaded ASAS $V$-band light curves and the accompanying catalog 
(ASAS Catalog of Variable Stars version 1.1; hereafter ACVS)
from the ASAS website 
(\href{http://www.astrouw.edu.pl/asas/?page=download}{http://www.astrouw.edu.pl/asas/?page=download})
\citep{Pojmanski1997AcA....47..467P}.
We selected only sources with a unique classification (e.g., DSCT), 
thereby excluding sources with multiple classifications (e.g., DSCT/RRC/EC). 
We also discarded MISC-type sources, which have unclear variability class.
After this filtering, the number of remaining sources is 10\,779 (Table \ref{tab:n_class_asas}).
The number of data points in the light curves varies between 200 and 2\,000, 
while 60\% of the light curves have more than 400 data points. 
The duration of the light curves is about 9 years.
The limiting magnitude of the ASAS survey is $V \sim 14$, which is
substantially brighter than the limits for OGLE, EROS-2, MACHO, and LINEAR.

Table \ref{tab:ASAS_testset} (left panel) shows the UPSILoN classification performance.
The average $F_1$ score is 0.85, demonstrating that most ASAS periodic variables are successfully classified.
However, $F_1$ scores are relatively low
for some classes such as RRL c and CEPH.
This could be associated with the poor quality
of the ACVS classification, which has been noted by previous studies also
\citep{Schmidt2009AJ....137.4598S, Richards2012ApJS..203...32R}.
For instance, \citet{Schmidt2009AJ....137.4598S} found that
there exist serious biases in the CEPH classification of ACVS.
\citet{Richards2012ApJS..203...32R}
found the significant discrepancies
between their classification and ACVS classification for RRL c and DSCT.
We also find that our low precision for DSCT is mainly caused
by the UPSILoN's misclassification of EB EC (contact binaries) as DSCT.
We visually checked the variability pattern of EB EC
and DSCT classified by ACVS, and found that they are generally similar,
as shown in Fig. \ref{fig:ASAS_EB_DSCT}.
The low precision of EB ESD (semi-detached binary), 0.57,
is likewise caused primarily by the confusion with EB EC, again
due to their similar variability patterns.
In contrast, the precision and recall of EB ED (detached binaries) 
are both fairly high: 0.93 and 0.96 respectively.
This is because the variability pattern of EB ED --
typically sharp flux gradients near its eclipses -- is
substantially different from that of EB EC or DSCT.

\section{Investigation of Classification Performance
in the Presence of Other Variability Types: 
the Hipparcos Periodic Variable Stars}
\label{sec:application_2}

We obtained a catalog of Hipparcos periodic variable stars
and their corresponding light curves\footnote{Periodic 
variables in Table 1 from \citet{Dubath2011MNRAS.414.2602D}.} 
from \citet{Dubath2011MNRAS.414.2602D}.
This dataset contains many  variability types 
(e.g., rotational variability, eruptive variability) 
that are not  in the UPSILoN training set, as well as types which are.
Thus this dataset is suitable for examining the potential contamination
to the UPSILoN classification by unaccounted for variability types.
We refined each Hipparcos  light curve by 3$\sigma$ clipping,
and by excluding light curves with spurious periods (see section \ref{sec:classification_model})
as we did for the MACHO, LINEAR, and ASAS datasets (see
section \ref{sec:application}).
Table \ref{tab:n_class_hip} gives the number of each Hipparcos variability type in the dataset
after this refinement. The table also shows the corresponding UPSILoN classes.
We then extracted the 16 features to predict a variability class.
The average number of data points in the light curves is about 100,
and the duration of the light curves varies from one to three years,
which is much fewer and shorter than the other three datasets.

\renewcommand{\arraystretch}{1.2}
\begin{table*}
\small
\begin{center}
\caption{The number of objects per true class in the Hipparcos dataset \label{tab:n_class_hip}}
\begin{tabular}{lrrr}
\hline\hline
Type & Hipparcos Class & Number & UPSILoN class \\
\hline
$\delta$ Scuti & DSCT      &  45 & DSCT \\
\,\,\, low amplitude & DSCTC   &    80 & DSCT \\
RR Lyrae & RRAB      &  71 & RRL ab\\
& RRC   &      19  &  RRL c \\
$\delta$ Cepheid & & \\
\,\,\, fundamental mode & DCEP   &    170  & CEPH F \\
\,\,\, first overtone & DCEPS &      28  &  CEPH 1O \\
\,\,\, multi-mode & CEP(B)  &    11 & CEPH Other \\
Eclipsing binary  & & \\
\,\,\,  detached & EA    &     222 &  EB ED \\
\,\,\,  semi-detached & EB    &     246  &  EB ESD \\
\,\,\,  contact & EW       &   99  & EB EC \\
Long-period variable & LPV    &    254 & LPV \\
Ellipsoidal & ELL     &    27 & \\
Slowly pulsating B star & SPB      &   78  & \\
$\alpha^2$ Canum Venaticorum & ACV      &   75  & \\
RS Canum Venaticorum & RS+BY     &  35  & \\
\,\,\, + BY Draconis & & & \\
$\gamma$ Doradus & GDOR       & 25  & \\
$\beta$ Cephei & BCEP      &  29  & \\
$\alpha$ Cygni & ACYG   &     17  & \\
B emmission line star & BE+GCAS &    12  & \\
\,\,\, + $\gamma$ Cassiopeiae & & & \\
SX Arietis & SXARI    &    7  & \\
W Virginis & & & \\
\,\,\, long-period (> 8 days) & CWA      &    9  & \\
\,\,\, short period  (< 8 days) & CWB    &      6  & \\
RV Tauri & RV        &   5  & \\
\hline
Total & & 1570 &  \\
\end{tabular}
\end{center}
\end{table*}

\begin{figure*}
\begin{center}
       \includegraphics[width=1.0\textwidth]{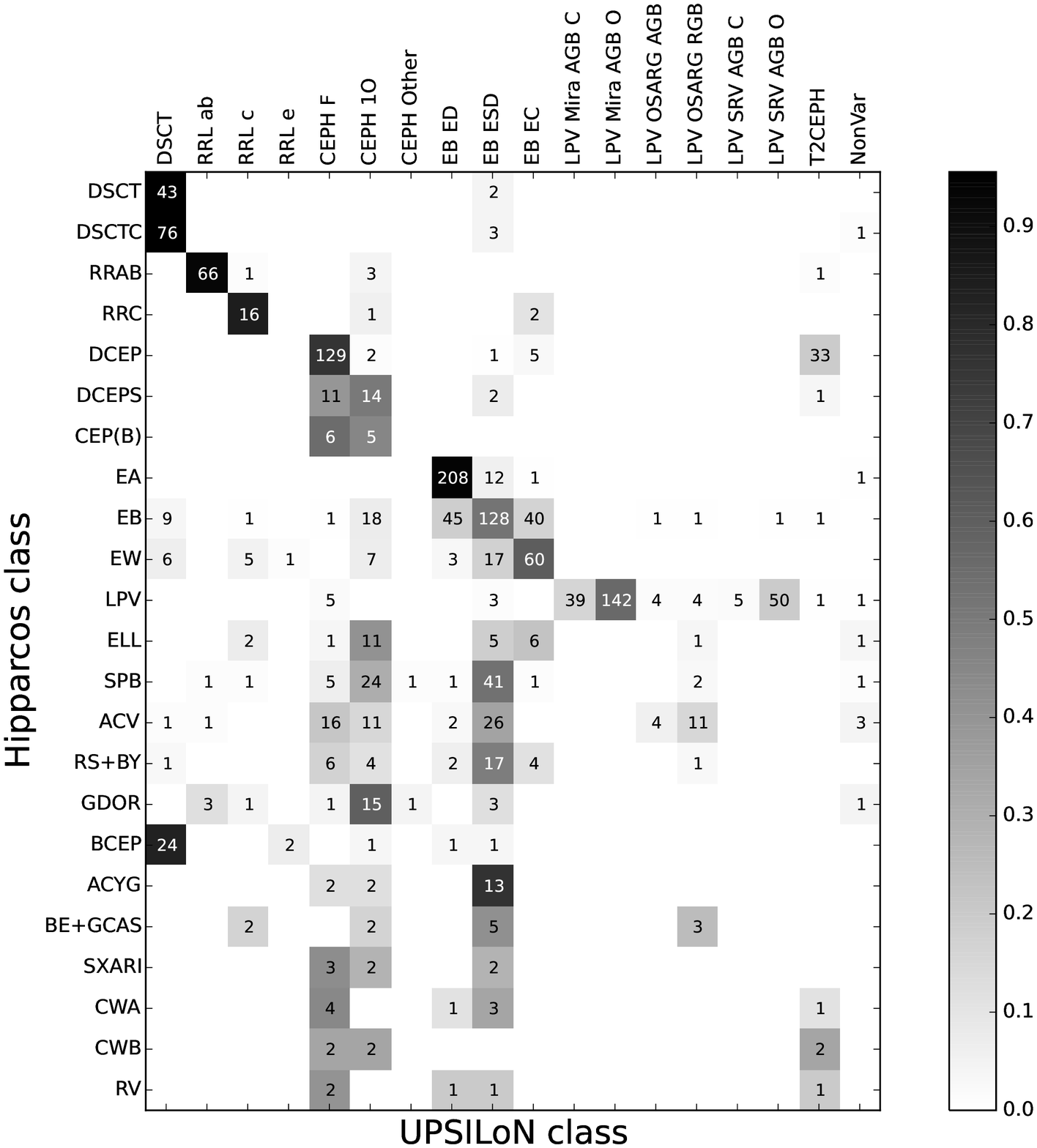}
\end{center}
    \caption{Confusion matrix of the Hipparcos dataset classified by the UPSILoN classifier.
Each cell-color represents the fraction of objects of that Hipparcos class (rows) assigned to the predicted UPSILoN classes (columns) on a gray scale. The value in each cell is the number of the Hipparcos sources.}
    \label{fig:hip_confusion_matrix}
\end{figure*}

In Fig. \ref{fig:hip_confusion_matrix}, we show the confusion matrix of the classification results.
This is more informative than a table of recall/precision 
since many Hipparcos classes do not exist in the training set.
As the matrix shows, the Hipparcos classes from DSCT to LPV 
(from top to middle row), which are those which correspond
directly to UPSILoN classes, 
are relatively classified well by the UPSILoN classifier.
The recall rates for those classes are generally high, up to 0.96, which indicates that
UPSILoN successfully classifies the major types of periodic variables
even though the number of data points is relatively small
and the duration is relatively short.
Nevertheless, we see the confusion between subclasses of eclipsing binaries,
which is expected following what we showed ealrier for the UPSILoN classifier performance 
(see Fig. \ref{fig:model_map_subclass}).
About 40\% of Hipparcos DCEPS (i.e., first overtone Cepheid)
are misclassified as CEPH F (i.e., fundamental mode Cepheid).
\citet{Dubath2011MNRAS.414.2602D} also found 
a similar degree of confusion between these two classes
and suggested that an overlap between the two classes in the period-amplitude space
might be the reason.
In addition, about half of the CEP(B) (i.e., multi-mode Cepheid) are classified as CEPH F
rather than CEPH Other. The rest are classified as CEPH 1O (first overtone).
In other words, all of the Hipparcos multi-mode Cepheid variables are misclassified.
This confusion could be associated with their multi-period characteristics,
because the UPSILoN classifier is trained using only a single period
but not using multiple periods, as also suggested by \citet{Dubath2011MNRAS.414.2602D}.
However this explanation might not clarify the confusion of this case
because the recall/precision of CEPH Other is relatively high, as 
shown in Fig. \ref{fig:model_map_subclass}.
Thus we suspect that the training set might
have an insufficient number of sources of the same variability type CEP(B).
Most of the training sources in the CEPH Other class are 
double-mode Cepheids F/1O and 1O/2O, whereas we do not know 
which pulsation modes these 11 CEP(B) stars are in.
Further investigation is therefore not possible.

The other Hipparcos classes, from ELL to RV, contaminate the classification results, 
which is not very surprising because the UPSILoN training set 
that defines its known class space does not contain those variability types.
The most contaminated UPSILoN classes are
DSCT, CEPH F, CEPH 1O, and EB ESD whereas other UPSILoN classes
are less contaminated or almost entirely uncontaminated by them
(i.e., all RRL subclasses, CEPH Other, EB ED, EB EC, all LPV subclasses, and T2CEPH).
As the table shows, 24 of the 29 BCEP are misclassified as DSCT due to 
BCEP's short periods and low amplitudes, which are also the characteristics of DSCT.
In addition, contamination caused by SPB, ACV, RS+BY, GDOR, and ACYG
to CEPH F, CEPH 1O and/or EB ESD is also high.
For instance, 90\% of SPB and 70\% of ACV are classified as one of these UPSILoN classes.
This might be because we did not use colors as a variability feature
to train the UPSILoN classifier, which is one of the most useful
features to separate different variability types. Training and validating an another model 
with survey-specific features (e.g., colors) is out of the scope of this paper.
Nevertheless, most variable star surveys, being area and magnitude limited,
will generally have a small fraction of variable stars
of the rare ``contaminating'' classes considered here.
Thus they would not be a significant source of contamination
for the UPSILoN classes. Of course, UPSILoN
will be unable to classify these rare classes, 
because it is not designed to do that.

\begin{figure*}
\begin{center}
\begin{minipage}[c]{9cm}
        \includegraphics[width=1.0\textwidth]{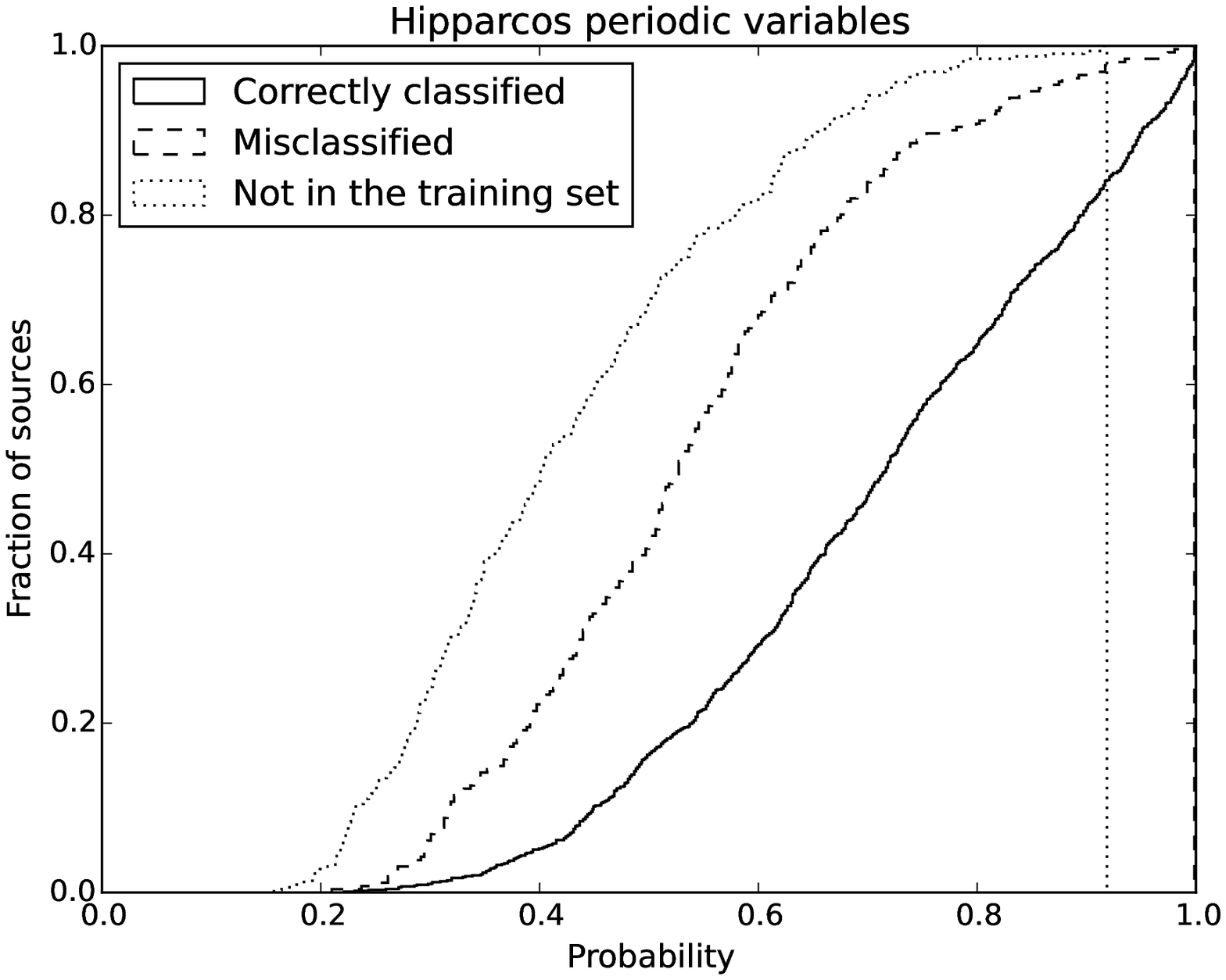}
\end{minipage}
\begin{minipage}[c]{9cm}
        \includegraphics[width=1.0\textwidth]{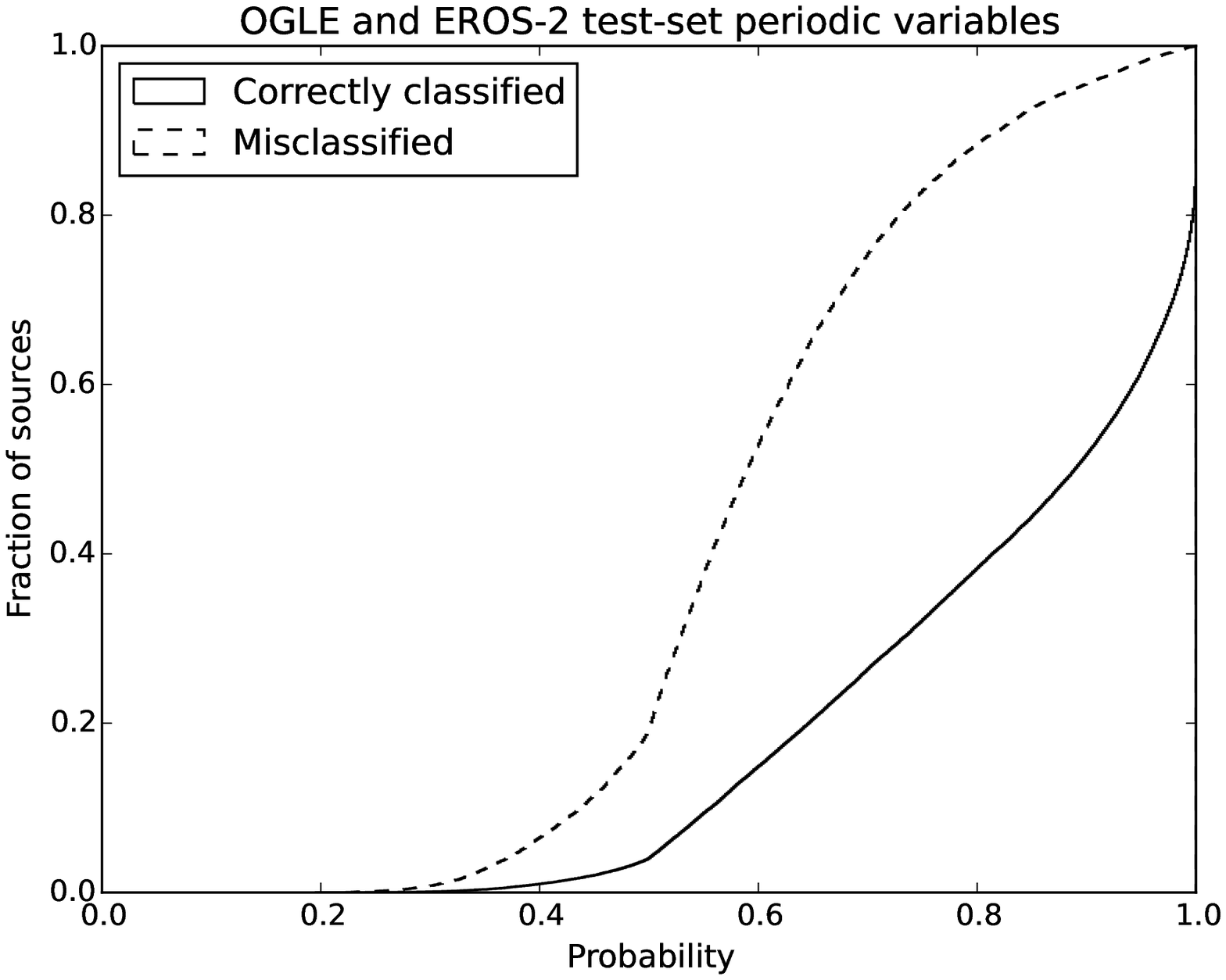}
\end{minipage}
\end{center}
    \caption
           {Fraction of sources (y-axis) whose probabilities of the accepted classes
           are lower than a certain probability (x-axis) 
           for the Hipparcos periodic variables (left panel) and
           the OGLE/EROS-2 validation test-set, $T_2$ (see section \ref{sec:training_classification_model}), (right panel).
           The correctly classified sources (solid line) have generally higher probability 
           than other two groups, each of which is
           1) the misclassified sources (dashed line), and 
           2) the sources whose variability types are not in the training set (dotted line).}
    \label{fig:Hipparcos_prob_hist}
\end{figure*}

Fig. \ref{fig:Hipparcos_prob_hist} shows the cumulative distribution
of the assigned class probabilities
for both the Hipparcos periodic variables (left panel)
and the OGLE/EROS-2 test-set periodic variables ($T_2$ from section \ref{sec:training_classification_model}, right panel)\footnote{Note that
the probabilities for the OGLE/EROS-2 test-set
were derived during the validation phase
using the model trained on only $T_1$
(section \ref{sec:training_classification_model}).}.
The solid line shows the sources
whose variable types are in the training set
and which are correctly classified by UPSILoN.
The dashed line shows the sources
of the same variability types but which are misclassified.
As both panels show, the correctly classified sources
tend to have higher probabilities than the misclassified sources.
More importantly, the sources whose variability types are not in 
the training set (dotted line) generally have lower probabilities 
than other two groups (solid line and dashed line).
This is not surprising, because UPSILoN does not know the data space which these variability types occupy. The data are poorly explained by several classes, no class dominates, so UPSILoN assigns all of them low probabilities (as the probabilities must sum to one).
Thus, in order to reduce the contamination, one can set a probability cut.
For instance, excluding sources with probabilities lower than 0.5
removes about 70\% of the contaminating sources and also
$\sim$40\% of the misclassified sources,
as can be seen in the left panel of Fig. \ref{fig:Hipparcos_prob_hist}.
Such cuts, however, will reduce the recall rate at the same time
(16\% loss of the correctly classified sources in this case).

\section{Investigation of Classifier Performance as a Function of Light Curve Length and Sampling}
\label{sec:resampled_test}

\begin{figure*}
\begin{center}
        \includegraphics[width=0.8\textwidth]{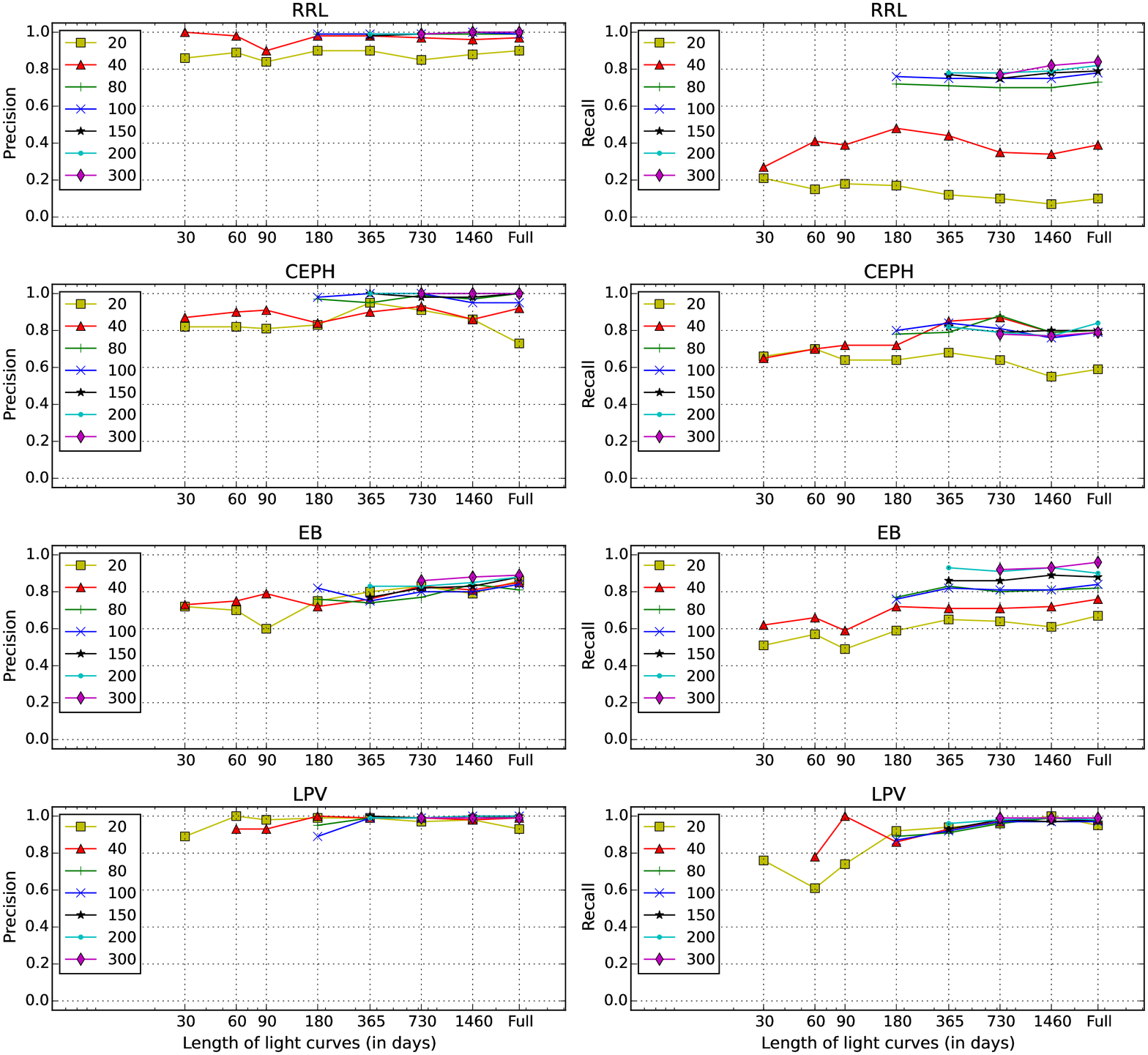}
\end{center}
    \caption
           {Precision (left column) and recall (right column) of the resampled MACHO dataset
           as a function of length of light curves, $l$ (horizontal axis),
           and the number of data points, $n$ (different lines), for different variability classes (rows).}
    \label{fig:quality_resampled_MACHO}
\end{figure*}

\begin{figure*}
\begin{center}
        \includegraphics[width=0.8\textwidth]{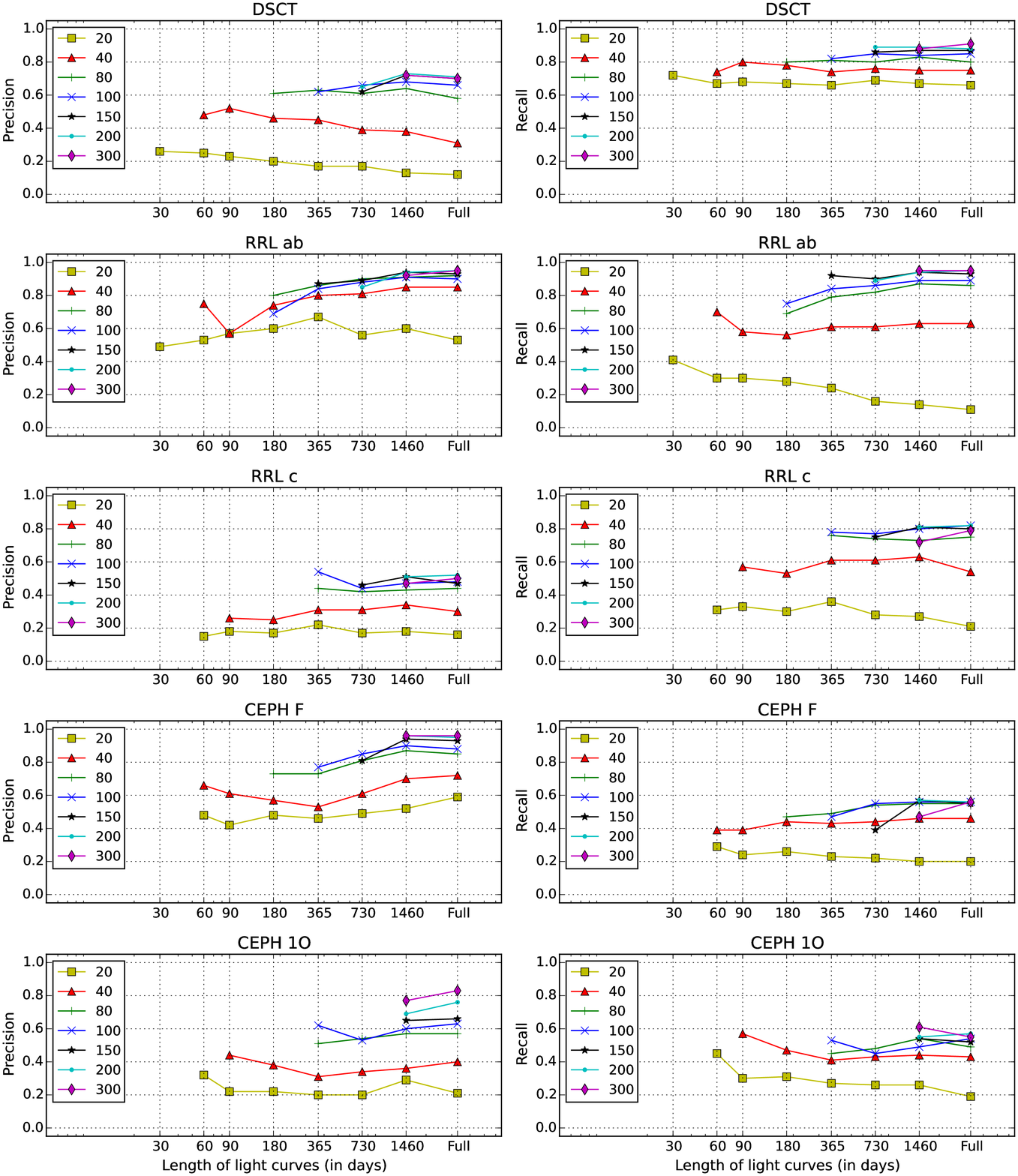}
\end{center}
    \caption
           {As Figure \ref{fig:quality_resampled_MACHO}, but for the resampled ASAS light curves.}
    \label{fig:quality_resampled_ASAS_1}
\end{figure*}

\begin{figure*}
\begin{center}
            \includegraphics[width=0.8\textwidth]{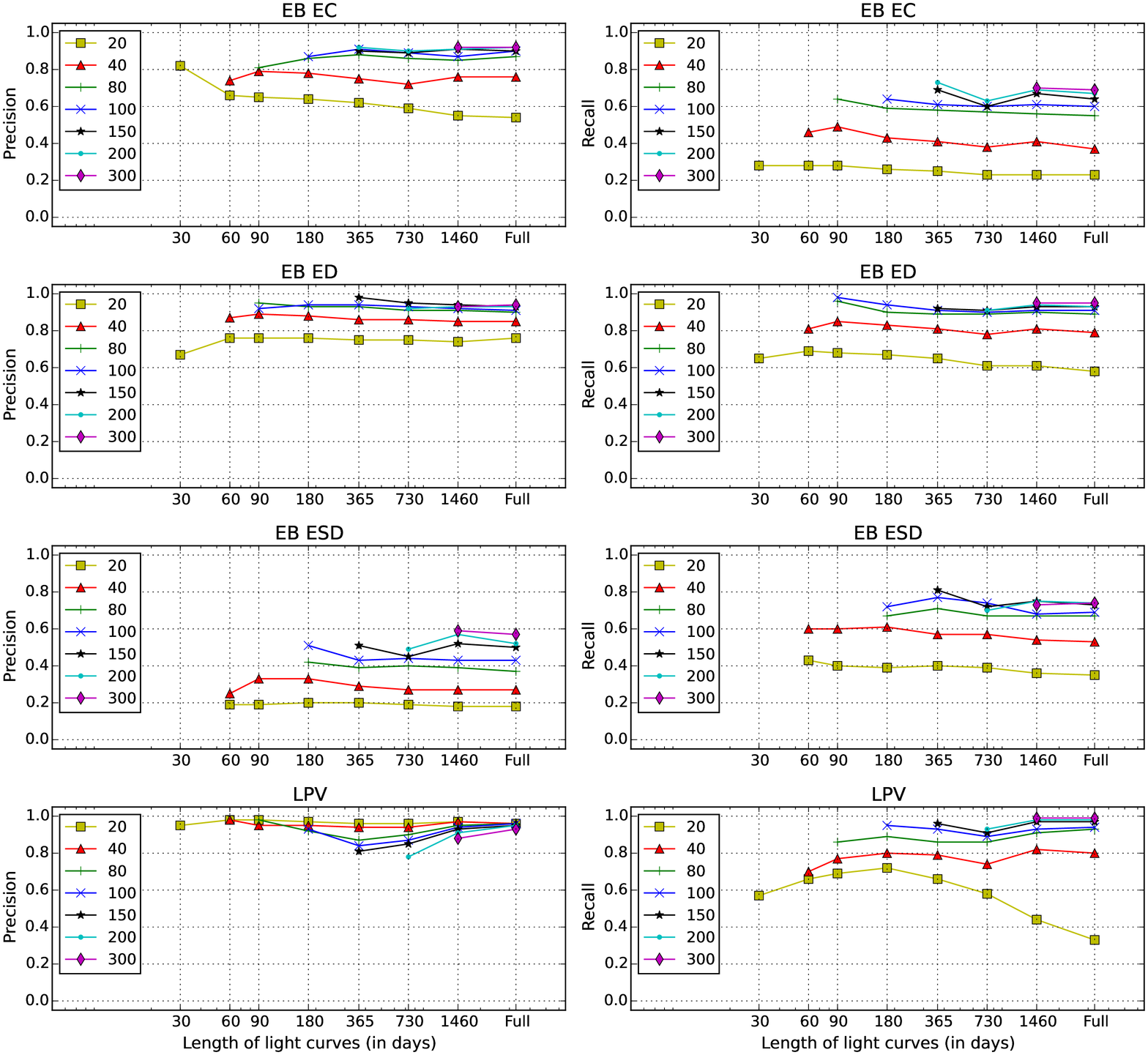}
\end{center}
    \caption
           {Continuation of Figure \ref{fig:quality_resampled_ASAS_1}.}
	\label{fig:quality_resampled_ASAS_2}
\end{figure*}

In this section we present the results of experiments
to assess the classification performance of the UPSILoN classifier
for light curves having different numbers of data points
and different observation durations.
To achieve this we constructed a set of light curves
by resampling the MACHO and ASAS light curves 
from the previous sections based on two parameters:
the length of a light curve, $l$ (i.e., observation duration);
and the number of data points, $n$.
We chose $l = $ 30, 60, 90, 180, 365, 730, and 1460 days, 
and also the full observation duration (i.e., 7 year for MACHO and 9 year for ASAS).
For the number of data points, 
we chose $n = $ 20, 40, 80, 100, 150, 200, and 300.
The resampling procedures were as follows. For a given light curve we

\begin{enumerate}

\item extract measurements observed between
the starting epoch of the light curve and $l$ days after the starting epoch.
If the number of the selected measurements is fewer than $n$, skip
the current light curve and move to the next light curve.

\item randomly select $n$ unique samples among these measurements. 

\end{enumerate}

{\noindent}These $n$ unique samples give a resampled light curve of duration $l$.
We did not resample the LINEAR and Hipparcos light curves
because most of them have fewer than 300 data points.

We applied the UPSILoN classifier to these resampled MACHO and ASAS light curves.  Fig. \ref{fig:quality_resampled_MACHO} shows the precision and recall for each variability class for the MACHO data.  Figs. \ref{fig:quality_resampled_ASAS_1} and \ref{fig:quality_resampled_ASAS_2} show the same for ASAS.  The horizontal-axis is the length of light curves, $l$, the vertical-axis is either precision or recall, and each line represents a different number of data points, $n$.  In general, the more data points, the higher the precision and recall.

In the case of MACHO (Fig. \ref{fig:quality_resampled_MACHO}),
precision rapidly reaches its maximum value
as the number of data points, $n$, reaches 80.
Even when there are fewer data points (e.g., 20 and 40),  precision is fairly high.
Length, $l$, hardly affects precision.
In contrast, recall is low for some classes (e.g., RRL)
when $n$ = 20 or 40. The reason for the low recall is because lots
of light curves were classified as non-variables.
This happens because the classification of
variables with relatively short periods, such as RRL and CEPH,
needs a sufficient number of data points over a short timescale.
As the figure shows, the recall of RRL and CEPH slightly decreases
as $l$ increases when $n$  = 20 or 40, for the same reason.
In addition, to quantify EB-like variability, 
which has steep flux gradients around the eclipses,
light curves have to be very well sampled to cover the
area around the flux gradients. Therefore, recall for EB is also relatively low.

Figs. \ref{fig:quality_resampled_ASAS_1}
and \ref{fig:quality_resampled_ASAS_2}
show the classification results of
the resampled ASAS light curves.
Precision and recall is generally lower than
MACHO's because
ACVS provides subclasses whereas MACHO does not. We saw earlier that recovering
subclasses is a harder problem than recovering superclasses, as we would expect.
The poor quality of the ACVS classification could also
contribute to the low recall and precision, as explained in 
the previous section.
Nonetheless, when there are more than 40 data points,
precision and recall are reasonably stable and rapidly reaches their maxima.
We also found that recall and precision of most of the classes, namely
DSCT, RRL ab, RRL c, CEPH, EB, and LPV,
decreases as $l$ increases when $n$  = 20 or 40,
as we saw from the resampled MACHO classification results.

From the experiments presented above,
we see that if the number of data points, $n$,
is larger than or equal to 80, precision and recall are not significantly
lower than what we can achieve using 
more data points and longer duration light curves.
More importantly, the duration of the light curves
does not significantly alter the classification quality if $n \geq 80$.

\section{Summary}
\label{sec:summary}

The quality of any supervised classification model depends strongly on the quality of the training data,
and yet constructing a clean and reasonably complete training set is quite difficult.
This is particularly the case in the early stages of a survey, 
when only a few known objects will have been observed, making it hard to build a classifier.

In this work, we have developed a general purpose classifier 
for periodic variable stars to help alleviate this problem.
First of all, it is known that periodic variables are
relatively straightforward to separate from non-periodic variables or non-variables
because of the clear periodic patterns of their light curves.
Second, lots of well-sampled light curves
from different surveys are available that can be used as a training set.
As the characteristics of periodic variables can be captured by
survey-independent features
(e.g., periods, amplitudes, shape of phase-folded light curves),
it is possible to build a universal training set.

We constructed a training set using the OGLE and EROS-2 periodic variables
consisting of 143\,923 well-sampled light curves
for a number of different types of periodic variable stars.
We mixed two different surveys having different characteristics
(sampling rates, observation durations, optical bands, etc.)
in order to train a classification model that is relatively tolerant of the diverse 
characteristics that could be encountered in the application datasets from other surveys.
We extracted 16 variability features that are
relatively survey-independent and used these features
to train a random forest classifier, the UPSILoN classifier.
Using periodic variables identified in the MACHO, LINEAR, and ASAS surveys,
we have shown that UPSILoN can successfully classify
periodic variable types in the training set with high precision and recall.
Using the Hipparcos periodic variables,
we found that the UPSILoN classes, CEPH F, CEPH 1O, and EB ESD,
can be contaminated by other variability types that are not in the training set.
We also tested the classifier using the resampled MACHO and ASAS
light curves, where we found that
precision and recall were fairly stable regardless of the time sampling and duration,
as long as there was a sufficient number of data points ($\geq$ 80).
The UPSILoN classifier should prove useful
for performing an initial classification into the major periodic
variability classes of light curves coming from a broad set of optical
monitoring surveys.

\section*{Acknowledgment}

We thank P. Dubath and L. Rimoldini for providing the Hipparcos periodic variable stars
and its catalog from their previous work  \citep{Dubath2011MNRAS.414.2602D}.
The analysis in this work has been done using the multi-core machine, \texttt{Mekong},
equipped with eight AMD Opteron 6276 CPUs
and 264 GB memory, at the Max-Planck Institute for Astronomy.

\bibliographystyle{aa}
\bibliography{bib}{}

\begin{thebibliography}{55}
\expandafter\ifx\csname natexlab\endcsname\relax\def\natexlab#1{#1}\fi

\bibitem[{{Alcock} {et~al.}(1996{\natexlab{a}}){Alcock}, {Allsman}, {Axelrod},
  {Bennett}, {Cook}, {Freeman}, {Griest}, {Guern}, {Lehner}, {Marshall},
  {Park}, {Perlmutter}, {Peterson}, {Pratt}, {Quinn}, {Rodgers}, {Stubbs}, \&
  {Sutherland}}]{Alcock1996ApJ...461...84A}
{Alcock}, C., {Allsman}, R.~A., {Axelrod}, T.~S., {et~al.} 1996{\natexlab{a}},
  ApJ, 461, 84

\bibitem[{{Alcock} {et~al.}(1996{\natexlab{b}}){Alcock}, {Allsman}, {Axelrod},
  {Bennett}, {Cook}, {Freeman}, {Griest}, {Marshall}, {Peterson}, {Pratt},
  {Quinn}, {Rodgers}, {Stubbs}, {Sutherland}, \&
  {Welch}}]{Alcock1996AJ....111.1146A}
{Alcock}, C., {Allsman}, R.~A., {Axelrod}, T.~S., {et~al.} 1996{\natexlab{b}},
  AJ, 111, 1146

\bibitem[{{Auvergne} {et~al.}(2009){Auvergne}, {Bodin}, {Boisnard}, {Buey},
  {Chaintreuil}, {Epstein}, {Jouret}, {Lam-Trong}, {Levacher}, {Magnan},
  {Perez}, {Plasson}, {Plesseria}, {Peter}, {Steller}, {Tiph{\`e}ne}, {Baglin},
  {Agogu{\'e}}, {Appourchaux}, {Barbet}, {Beaufort}, {Bellenger}, {Berlin},
  {Bernardi}, {Blouin}, {Boumier}, {Bonneau}, {Briet}, {Butler}, {Cautain},
  {Chiavassa}, {Costes}, {Cuvilho}, {Cunha-Parro}, {de Oliveira Fialho},
  {Decaudin}, {Defise}, {Djalal}, {Docclo}, {Drummond}, {Dupuis}, {Exil},
  {Faur{\'e}}, {Gaboriaud}, {Gamet}, {Gavalda}, {Grolleau}, {Gueguen},
  {Guivarc'h}, {Guterman}, {Hasiba}, {Huntzinger}, {Hustaix}, {Imbert},
  {Jeanville}, {Johlander}, {Jorda}, {Journoud}, {Karioty}, {Kerjean},
  {Lafond}, {Lapeyrere}, {Landiech}, {Larqu{\'e}}, {Laudet}, {Le Merrer},
  {Leporati}, {Leruyet}, {Levieuge}, {Llebaria}, {Martin}, {Mazy}, {Mesnager},
  {Michel}, {Moalic}, {Monjoin}, {Naudet}, {Neukirchner}, {Nguyen-Kim},
  {Ollivier}, {Orcesi}, {Ottacher}, {Oulali}, {Parisot}, {Perruchot},
  {Piacentino}, {Pinheiro da Silva}, {Platzer}, {Pontet}, {Pradines},
  {Quentin}, {Rohbeck}, {Rolland}, {Rollenhagen}, {Romagnan}, {Russ}, {Samadi},
  {Schmidt}, {Schwartz}, {Sebbag}, {Smit}, {Sunter}, {Tello}, {Toulouse},
  {Ulmer}, {Vandermarcq}, {Vergnault}, {Wallner}, {Waultier}, \&
  {Zanatta}}]{Auvergne2009AA...506..411A}
{Auvergne}, M., {Bodin}, P., {Boisnard}, L., {et~al.} 2009, A\&A, 506, 411

\bibitem[{{Bailer-Jones} {et~al.}(2013){Bailer-Jones}, {Andrae}, {Arcay},
  {Astraatmadja}, {Bellas-Velidis}, {Berihuete}, {Bijaoui}, {Carri{\'o}n},
  {Dafonte}, {Damerdji}, {Dapergolas}, {de Laverny}, {Delchambre}, {Drazinos},
  {Drimmel}, {Fr{\'e}mat}, {Fustes}, {Garc{\'{\i}}a-Torres}, {Gu{\'e}d{\'e}},
  {Heiter}, {Janotto}, {Karampelas}, {Kim}, {Knude}, {Kolka}, {Kontizas},
  {Kontizas}, {Korn}, {Lanzafame}, {Lebreton}, {Lindstr{\o}m}, {Liu},
  {Livanou}, {Lobel}, {Manteiga}, {Martayan}, {Ordenovic}, {Pichon},
  {Recio-Blanco}, {Rocca-Volmerange}, {Sarro}, {Smith}, {Sordo}, {Soubiran},
  {Surdej}, {Th{\'e}venin}, {Tsalmantza}, {Vallenari}, \&
  {Zorec}}]{BailerJones2013AA...559A..74B}
{Bailer-Jones}, C.~A.~L., {Andrae}, R., {Arcay}, B., {et~al.} 2013, A\&A, 559,
  A74

\bibitem[{{Blomme} {et~al.}(2011){Blomme}, {Sarro}, {O'Donovan}, {Debosscher},
  {Brown}, {Lopez}, {Dubath}, {Rimoldini}, {Charbonneau}, {Dunham},
  {Mandushev}, {Ciardi}, {De Ridder}, \& {Aerts}}]{Blomme2011MNRAS.418...96B}
{Blomme}, J., {Sarro}, L.~M., {O'Donovan}, F.~T., {et~al.} 2011, MNRAS, 418, 96

\bibitem[{Breiman(2001)}]{Breiman2001}
Breiman, L. 2001, Machine Learning, 45, 5

\bibitem[{Breiman {et~al.}(1984)Breiman, Friedman, Stone, \&
  Olshen}]{breiman1984classification}
Breiman, L., Friedman, J., Stone, C., \& Olshen, R. 1984, Classification and
  Regression Trees, The Wadsworth and Brooks-Cole statistics-probability series
  (Taylor \& Francis)

\bibitem[{{Brown} \& {Gilliland}(1994)}]{Brown1994ARAA..32...37B}
{Brown}, T.~M. \& {Gilliland}, R.~L. 1994, ARA\&A, 32, 37

\bibitem[{{Carliles} {et~al.}(2010){Carliles}, {Budav{\'a}ri}, {Heinis},
  {Priebe}, \& {Szalay}}]{Carliles2010ApJ...712..511C}
{Carliles}, S., {Budav{\'a}ri}, T., {Heinis}, S., {Priebe}, C., \& {Szalay},
  A.~S. 2010, ApJ, 712, 511

\bibitem[{{Carretta} {et~al.}(2000){Carretta}, {Gratton}, {Clementini}, \&
  {Fusi Pecci}}]{Carretta2000ApJ...533..215C}
{Carretta}, E., {Gratton}, R.~G., {Clementini}, G., \& {Fusi Pecci}, F. 2000,
  ApJ, 533, 215

\bibitem[{{Catelan}(2009)}]{Catelan2009ApSS.320..261C}
{Catelan}, M. 2009, Astrophysics and Space Science, 320, 261

\bibitem[{{Debosscher} {et~al.}(2009){Debosscher}, {Sarro}, {L{\'o}pez},
  {Deleuil}, {Aerts}, {Auvergne}, {Baglin}, {Baudin}, {Chadid}, {Charpinet},
  {Cuypers}, {De Ridder}, {Garrido}, {Hubert}, {Janot-Pacheco}, {Jorda},
  {Kaiser}, {Kallinger}, {Kollath}, {Maceroni}, {Mathias}, {Michel}, {Moutou},
  {Neiner}, {Ollivier}, {Samadi}, {Solano}, {Surace}, {Vandenbussche}, \&
  {Weiss}}]{Debosscher2009AA...506..519D}
{Debosscher}, J., {Sarro}, L.~M., {L{\'o}pez}, M., {et~al.} 2009, A\&A, 506,
  519

\bibitem[{{Dubath} {et~al.}(2011){Dubath}, {Rimoldini}, {S{\"u}veges},
  {Blomme}, {L{\'o}pez}, {Sarro}, {De Ridder}, {Cuypers}, {Guy}, {Lecoeur},
  {Nienartowicz}, {Jan}, {Beck}, {Mowlavi}, {De Cat}, {Lebzelter}, \&
  {Eyer}}]{Dubath2011MNRAS.414.2602D}
{Dubath}, P., {Rimoldini}, L., {S{\"u}veges}, M., {et~al.} 2011, MNRAS, 414,
  2602

\bibitem[{{Eyer} {et~al.}(2014){Eyer}, {Evans}, {Mowlavi}, {Lanzafame},
  {Cuypers}, {De Ridder}, {Sarro}, {Clementini}, {Guy}, {Holl}, {Ordonez},
  {Nienartowicz}, \& {Lecoeur-Taibi}}]{Eyer2014EAS....67...75E}
{Eyer}, L., {Evans}, D.~W., {Mowlavi}, N., {et~al.} 2014, in EAS Publications
  Series, Vol.~67, EAS Publications Series, 75--78

\bibitem[{{Freedman} {et~al.}(2001){Freedman}, {Madore}, {Gibson}, {Ferrarese},
  {Kelson}, {Sakai}, {Mould}, {Kennicutt}, {Ford}, {Graham}, {Huchra},
  {Hughes}, {Illingworth}, {Macri}, \& {Stetson}}]{Freedman2001ApJ...553...47F}
{Freedman}, W.~L., {Madore}, B.~F., {Gibson}, B.~K., {et~al.} 2001, ApJ, 553,
  47

\bibitem[{{Graczyk} {et~al.}(2011){Graczyk}, {Soszy{\'n}ski}, {Poleski},
  {Pietrzy{\'n}ski}, {Udalski}, {Szyma{\'n}ski}, {Kubiak}, {Wyrzykowski}, \&
  {Ulaczyk}}]{Graczyk2011AcA....61..103G}
{Graczyk}, D., {Soszy{\'n}ski}, I., {Poleski}, R., {et~al.} 2011, Acta
  Astronomica, 61, 103

\bibitem[{{Graham} {et~al.}(2013){Graham}, {Drake}, {Djorgovski}, {Mahabal},
  {Donalek}, {Duan}, \& {Maker}}]{Graham2013MNRAS.434.3423G}
{Graham}, M.~J., {Drake}, A.~J., {Djorgovski}, S.~G., {et~al.} 2013, MNRAS,
  434, 3423

\bibitem[{Hastie {et~al.}(2009)Hastie, Tibshirani, \& Friedman}]{Hastie09}
Hastie, T., Tibshirani, R., \& Friedman, J. 2009, The elements of statistical
  learning: data mining, inference and prediction, 2nd edn. (Springer)

\bibitem[{{Ivezic} {et~al.}(2008){Ivezic}, {Tyson}, {Allsman}, {Andrew},
  {Angel}, \& {for the LSST Collaboration}}]{Ivezic2008arXiv0805.2366I}
{Ivezic}, Z., {Tyson}, J.~A., {Allsman}, R., {et~al.} 2008, ArXiv e-prints

\bibitem[{{Kim} {et~al.}(2014){Kim}, {Protopapas}, {Bailer-Jones}, {Byun},
  {Chang}, {Marquette}, \& {Shin}}]{Kim2014AA...566A..43K}
{Kim}, D.-W., {Protopapas}, P., {Bailer-Jones}, C.~A.~L., {et~al.} 2014, A\&A,
  566, A43

\bibitem[{{Kim} {et~al.}(2011){Kim}, {Protopapas}, {Byun}, {Alcock}, {Khardon},
  \& {Trichas}}]{Kim2011ApJ...735...68K}
{Kim}, D.-W., {Protopapas}, P., {Byun}, Y.-I., {et~al.} 2011, ApJ, 735, 68

\bibitem[{{Kim} {et~al.}(2012){Kim}, {Protopapas}, {Trichas}, {Rowan-Robinson},
  {Khardon}, {Alcock}, \& {Byun}}]{Kim2012ApJ...747..107K}
{Kim}, D.-W., {Protopapas}, P., {Trichas}, M., {et~al.} 2012, ApJ, 747, 107

\bibitem[{{Kunder} \& {Chaboyer}(2008)}]{Kunder2008AJ....136.2441K}
{Kunder}, A. \& {Chaboyer}, B. 2008, AJ, 136, 2441

\bibitem[{{Long} {et~al.}(2012){Long}, {Karoui}, {Rice}, {Richards}, \&
  {Bloom}}]{Long2012PASP..124..280L}
{Long}, J.~P., {Karoui}, N.~E., {Rice}, J.~A., {Richards}, J.~W., \& {Bloom},
  J.~S. 2012, PASP, 124, 280

\bibitem[{{Mackay}(2003)}]{Mackay2003itil.book.....M}
{Mackay}, D.~J.~C. 2003, {Information Theory, Inference and Learning
  Algorithms} (Cambridge University Press)

\bibitem[{{Masci} {et~al.}(2014){Masci}, {Hoffman}, {Grillmair}, \&
  {Cutri}}]{Masci2014AJ....148...21M}
{Masci}, F.~J., {Hoffman}, D.~I., {Grillmair}, C.~J., \& {Cutri}, R.~M. 2014,
  AJ, 148, 21

\bibitem[{{Nun} {et~al.}(2014){Nun}, {Pichara}, {Protopapas}, \&
  {Kim}}]{Nun2014ApJ...793...23N}
{Nun}, I., {Pichara}, K., {Protopapas}, P., \& {Kim}, D.-W. 2014, ApJ, 793, 23

\bibitem[{{O'Donovan} {et~al.}(2009){O'Donovan}, {Charbonneau}, {Mandushev},
  {Dunham}, {Latham}, {Torres}, {Sozzetti}, {Brown}, {Trauger}, {Belmonte},
  {Rabus}, {Almenara}, {Alonso}, {Deeg}, {Esquerdo}, {Falco}, {Hillenbrand},
  {Roussanova}, {Stefanik}, \& {Winn}}]{ODonovan2009nsted.cat....6O}
{O'Donovan}, F.~T., {Charbonneau}, D., {Mandushev}, G., {et~al.} 2009, in
  NASA/IPAC/NExScI Star and Exoplanet Database, TrES Lyr1 Catalog, 6

\bibitem[{{Paegert} {et~al.}(2014){Paegert}, {Stassun}, \&
  {Burger}}]{Paegert2014AJ....148...31P}
{Paegert}, M., {Stassun}, K.~G., \& {Burger}, D.~M. 2014, AJ, 148, 31

\bibitem[{{Palaversa} {et~al.}(2013){Palaversa}, {Ivezi{\'c}}, {Eyer}, {Ru{\v
  z}djak}, {Sudar}, {Galin}, {Kroflin}, {Mesari{\'c}}, {Munk}, {Vrbanec},
  {Bo{\v z}i{\'c}}, {Loebman}, {Sesar}, {Rimoldini}, {Hunt-Walker},
  {VanderPlas}, {Westman}, {Stuart}, {Becker}, {Srdo{\v c}}, {Wozniak}, \&
  {Oluseyi}}]{Palaversa2013AJ....146..101P}
{Palaversa}, L., {Ivezi{\'c}}, {\v Z}., {Eyer}, L., {et~al.} 2013, AJ, 146, 101

\bibitem[{{Perryman} {et~al.}(2001){Perryman}, {de Boer}, {Gilmore}, {H{\o}g},
  {Lattanzi}, {Lindegren}, {Luri}, {Mignard}, {Pace}, \& {de
  Zeeuw}}]{Perryman2001AA...369..339P}
{Perryman}, M.~A.~C., {de Boer}, K.~S., {Gilmore}, G., {et~al.} 2001, A\&A,
  369, 339

\bibitem[{{Perryman} {et~al.}(1997){Perryman}, {Lindegren}, {Kovalevsky},
  {Hoeg}, {Bastian}, {Bernacca}, {Cr{\'e}z{\'e}}, {Donati}, {Grenon},
  {Grewing}, {van Leeuwen}, {van der Marel}, {Mignard}, {Murray}, {Le Poole},
  {Schrijver}, {Turon}, {Arenou}, {Froeschl{\'e}}, \&
  {Petersen}}]{Perryman1997AA...323L..49P}
{Perryman}, M.~A.~C., {Lindegren}, L., {Kovalevsky}, J., {et~al.} 1997, A\&A,
  323, L49

\bibitem[{{Petersen}(1986)}]{Petersen1986AA...170...59P}
{Petersen}, J.~O. 1986, A\&A, 170, 59

\bibitem[{{Pichara} {et~al.}(2012){Pichara}, {Protopapas}, {Kim}, {Marquette},
  \& {Tisserand}}]{Pichara2012MNRAS.427.1284P}
{Pichara}, K., {Protopapas}, P., {Kim}, D.-W., {Marquette}, J.-B., \&
  {Tisserand}, P. 2012, MNRAS, 427, 1284

\bibitem[{{Pojmanski}(1997)}]{Pojmanski1997AcA....47..467P}
{Pojmanski}, G. 1997, Acta Astronomica, 47, 467

\bibitem[{{Poleski} {et~al.}(2010){Poleski}, {Soszy{\~n}ski}, {Udalski},
  {Szyma{\~n}ski}, {Kubiak}, {Pietrzy{\~n}ski}, {Wyrzykowski}, {Szewczyk}, \&
  {Ulaczyk}}]{Poleski2010AcA....60....1P}
{Poleski}, R., {Soszy{\~n}ski}, I., {Udalski}, A., {et~al.} 2010, Acta
  Astronomica, 60, 1

\bibitem[{Quinlan(1993)}]{Quinlan1993}
Quinlan, J.~R. 1993, C4.5: programs for machine learning (San Francisco, CA,
  USA: Morgan Kaufmann Publishers Inc.)

\bibitem[{{Richards} {et~al.}(2012{\natexlab{a}}){Richards}, {Starr}, {Brink},
  {Miller}, {Bloom}, {Butler}, {James}, {Long}, \&
  {Rice}}]{Richards2012ApJ...744..192R}
{Richards}, J.~W., {Starr}, D.~L., {Brink}, H., {et~al.} 2012{\natexlab{a}},
  ApJ, 744, 192

\bibitem[{{Richards} {et~al.}(2012{\natexlab{b}}){Richards}, {Starr}, {Miller},
  {Bloom}, {Butler}, {Brink}, \& {Crellin-Quick}}]{Richards2012ApJS..203...32R}
{Richards}, J.~W., {Starr}, D.~L., {Miller}, A.~A., {et~al.}
  2012{\natexlab{b}}, ApJS, 203, 32

\bibitem[{{Riess} {et~al.}(2011){Riess}, {Macri}, {Casertano}, {Lampeitl},
  {Ferguson}, {Filippenko}, {Jha}, {Li}, \&
  {Chornock}}]{Riess2011ApJ...730..119R}
{Riess}, A.~G., {Macri}, L., {Casertano}, S., {et~al.} 2011, ApJ, 730, 119

\bibitem[{Saculinggan \& Balase(2013)}]{Saculinggan12013}
Saculinggan, M. \& Balase, E.~A. 2013, Journal of Physics: Conference Series,
  435, 012041

\bibitem[{{Schmidt} {et~al.}(2009){Schmidt}, {Hemen}, {Rogalla}, \&
  {Thacker-Lynn}}]{Schmidt2009AJ....137.4598S}
{Schmidt}, E.~G., {Hemen}, B., {Rogalla}, D., \& {Thacker-Lynn}, L. 2009, AJ,
  137, 4598

\bibitem[{{Sesar}(2011)}]{Sesar2011rrls.conf..135S}
{Sesar}, B. 2011, in RR Lyrae Stars, Metal-Poor Stars, and the Galaxy, ed.
  A.~{McWilliam}, 135

\bibitem[{Shapiro \& Wilk(1965)}]{shapiro1965analysis}
Shapiro, S.~S. \& Wilk, M.~B. 1965, Biometrika, 52, 591

\bibitem[{{Soszy{\'n}ski} {et~al.}(2008{\natexlab{a}}){Soszy{\'n}ski},
  {Poleski}, {Udalski}, {Szymanski}, {Kubiak}, {Pietrzynski}, {Wyrzykowski},
  {Szewczyk}, \& {Ulaczyk}}]{Soszynski2008AcA....58..163S}
{Soszy{\'n}ski}, I., {Poleski}, R., {Udalski}, A., {et~al.} 2008{\natexlab{a}},
  Acta Astronomica, 58, 163

\bibitem[{{Soszy{\'n}ski} {et~al.}(2009{\natexlab{a}}){Soszy{\'n}ski},
  {Udalski}, {Szyma{\~n}ski}, {Kubiak}, {Pietrzy{\~n}ski}, {Wyrzykowski},
  {Szewczyk}, {Ulaczyk}, \& {Poleski}}]{Soszynski2009AcA....59..239S}
{Soszy{\'n}ski}, I., {Udalski}, A., {Szyma{\~n}ski}, M.~K., {et~al.}
  2009{\natexlab{a}}, Acta Astronomica, 59, 239

\bibitem[{{Soszy{\'n}ski} {et~al.}(2008{\natexlab{b}}){Soszy{\'n}ski},
  {Udalski}, {Szyma{\'n}ski}, {Kubiak}, {Pietrzy{\'n}ski}, {Wyrzykowski},
  {Szewczyk}, {Ulaczyk}, \& {Poleski}}]{Soszynski2008AcA....58..293S}
{Soszy{\'n}ski}, I., {Udalski}, A., {Szyma{\'n}ski}, M.~K., {et~al.}
  2008{\natexlab{b}}, Acta Astronomica, 58, 293

\bibitem[{{Soszy{\'n}ski} {et~al.}(2009{\natexlab{b}}){Soszy{\'n}ski},
  {Udalski}, {Szyma{\'n}ski}, {Kubiak}, {Pietrzy{\'n}ski}, {Wyrzykowski},
  {Szewczyk}, {Ulaczyk}, \& {Poleski}}]{Soszynski2009AcA....59....1S}
{Soszy{\'n}ski}, I., {Udalski}, A., {Szyma{\'n}ski}, M.~K., {et~al.}
  2009{\natexlab{b}}, Acta Astronomica, 59, 1

\bibitem[{{Stokes} {et~al.}(2000){Stokes}, {Evans}, {Viggh}, {Shelly}, \&
  {Pearce}}]{Stokes2000Icar..148...21S}
{Stokes}, G.~H., {Evans}, J.~B., {Viggh}, H.~E.~M., {Shelly}, F.~C., \&
  {Pearce}, E.~C. 2000, Icarus, 148, 21

\bibitem[{{Szymanski}(2005)}]{Szymanski2005AcA....55...43S}
{Szymanski}, M.~K. 2005, Acta Astronomica, 55, 43

\bibitem[{{Tisserand} {et~al.}(2007){Tisserand}, {Le Guillou}, {Afonso},
  {Albert}, {Andersen}, {Ansari}, {Aubourg}, {Bareyre}, {Beaulieu}, {Charlot},
  {Coutures}, {Ferlet}, {Fouqu{\'e}}, {Glicenstein}, {Goldman}, {Gould},
  {Graff}, {Gros}, {Haissinski}, {Hamadache}, {de Kat}, {Lasserre}, {Lesquoy},
  {Loup}, {Magneville}, {Marquette}, {Maurice}, {Maury}, {Milsztajn}, {Moniez},
  {Palanque-Delabrouille}, {Perdereau}, {Rahal}, {Rich}, {Spiro},
  {Vidal-Madjar}, {Vigroux}, {Zylberajch}, \& {EROS-2
  Collaboration}}]{Tisserand2007AA...469..387T}
{Tisserand}, P., {Le Guillou}, L., {Afonso}, C., {et~al.} 2007, A\&A, 469, 387

\bibitem[{{Torres} {et~al.}(2010){Torres}, {Andersen}, \&
  {Gim{\'e}nez}}]{Torres2010AARv..18...67T}
{Torres}, G., {Andersen}, J., \& {Gim{\'e}nez}, A. 2010, The Astronomy and
  Astrophysics Review, 18, 67

\bibitem[{{Udalski} {et~al.}(1997){Udalski}, {Kubiak}, \&
  {Szymanski}}]{Udalski1997AcA....47..319U}
{Udalski}, A., {Kubiak}, M., \& {Szymanski}, M. 1997, Acta Astronomica, 47, 319

\bibitem[{{Vivas} {et~al.}(2001){Vivas}, {Zinn}, {Andrews}, {Bailyn}, {Baltay},
  {Coppi}, {Ellman}, {Girard}, {Rabinowitz}, {Schaefer}, {Shin}, {Snyder},
  {Sofia}, {van Altena}, {Abad}, {Bongiovanni}, {Brice{\~n}o}, {Bruzual},
  {Della Prugna}, {Herrera}, {Magris}, {Mateu}, {Pacheco}, {S{\'a}nchez},
  {S{\'a}nchez}, {Schenner}, {Stock}, {Vicente}, {Vieira}, {Ferr{\'{\i}}n},
  {Hernandez}, {Gebhard}, {Honeycutt}, {Mufson}, {Musser}, \&
  {Rengstorf}}]{Vivas2001ApJ...554L..33V}
{Vivas}, A.~K., {Zinn}, R., {Andrews}, P., {et~al.} 2001, ApJ, 554, L33

\bibitem[{{von Neumann}(1941)}]{Neumann1941}
{von Neumann}, J. 1941, Ann. Math. Statist., 12, 367

\end{thebibliography}

\section*{Appendix}

\definecolor{mygreen}{rgb}{0,0.6,0}
\definecolor{mygray}{rgb}{0.5,0.5,0.5}
\definecolor{mymauve}{rgb}{0.58,0,0.82}

\lstset{ %
  backgroundcolor=\color{white},   % choose the background color; you must add \usepackage{color} or \usepackage{xcolor}
  basicstyle=\footnotesize,        % the size of the fonts that are used for the code
  breakatwhitespace=false,         % sets if automatic breaks should only happen at whitespace
  breaklines=true,                 % sets automatic line breaking
  captionpos=b,                    % sets the caption-position to bottom
  commentstyle=\color{mygray},    % comment style
  deletekeywords={...},            % if you want to delete keywords from the given language
  escapeinside={\%*}{*)},          % if you want to add LaTeX within your code
  extendedchars=true,              % lets you use non-ASCII characters; for 8-bits encodings only, does not work with UTF-8
  frame=single,                    % adds a frame around the code
  keepspaces=true,                 % keeps spaces in text, useful for keeping indentation of code (possibly needs columns=flexible)
  keywordstyle=\color{blue}\bf,       % keyword style
  language=Python,                 % the language of the code
  otherkeywords={*},            % if you want to add more keywords to the set
  numbers=none,                    % where to put the line-numbers; possible values are (none, left, right)
  numbersep=8pt,                   % how far the line-numbers are from the code
  numberstyle=\tiny\color{mygreen}, % the style that is used for the line-numbers
  rulecolor=\color{black},         % if not set, the frame-color may be changed on line-breaks within not-black text (e.g. comments (green here))
  showspaces=false,                % show spaces everywhere adding particular underscores; it overrides 'showstringspaces'
  showstringspaces=false,          % underline spaces within strings only
  showtabs=false,                  % show tabs within strings adding particular underscores
  stepnumber=2,                    % the step between two line-numbers. If it's 1, each line will be numbered
  stringstyle=\color{mymauve},     % string literal style
  tabsize=1,                       % sets default tabsize to 2 spaces
  title=\lstname                   % show the filename of files included with \lstinputlisting; also try caption instead of title
}

\subsection*{How to Use UPSILoN}

The following pseudo-code shows how to use the UPSILoN library
to extract features from light curves and then to classify them.
        
\scriptsize
\begin{lstlisting}[language=Python, frame=single]
# Import the UPSILoN library.
import upsilon

# Load a random forest model.
rf_model = upsilon.load_rf_model()

# Extract features and classify. 
for light_curve in set_of_light_curves:

    # Read a light curve.
    date = np.array([:])
    mag = np.array([:])
    err = np.array([:])
    
    # Extract features.
    e_fts = upsilon.ExtractFeatures(date, mag, err)
    e_fts.run()
    features = e_fts.get_features()
    
    # Classify the light curve.
    label, probability, flag = upsilon.predict(rf_model, features)
\end{lstlisting}
\normalsize

{\noindent}\texttt{label} and \texttt{probability}
are a predicted class and a class probability, respectively.
The \texttt{flag} gives additional information of the classification result.
For details of how to use the UPSILoN library,
visit the \href{https://goo.gl/xmFO6Q}{GitHub repository (https://goo.gl/xmFO6Q)}.
UPSILoN is released under the MIT license, so it is free to use, copy, and modify.

If you use UPSILoN in scientific publication, we would appreciate citations to this paper.

\subsection*{UPSILoN Runtime}

We used the Macbook Pro 13$^{\prime\prime}$ equipped with 2.7 GHz Intel Core i5, 8 GB memory, and 256 GB SSD to measure the UPSILoN runtime\footnote{Wall-clock time.}
using the EROS-2 light curves, which contains about 500 data points.
The average runtime for feature extraction per light curve is $\sim$1.1 seconds.
The most time-consuming part of the feature extraction is the period estimation. 
UPSILoN can reduce the period estimation runtime 
by utilizing multiple cores to speed up the Fourier transform part in the Lomb-Scargle algorithm.
We observed that the period estimation runtime is inversely scaled with the number of cores.
The runtime for class prediction per light curve is less than 0.01 seconds,
which is negligible when compared to the feature extraction runtime.
In addition, loading the UPSILoN random forest model 
file takes $\sim$4 seconds.
Note that the loading is not a recurring task.

\end{document}